\documentclass[aps,prl,reprint,superscriptaddress]{revtex4-2}
\usepackage{diagbox}
\usepackage{upgreek}
\usepackage{amsmath}
\usepackage{amsfonts}
\usepackage{physics}
\usepackage[space]{grffile}
\usepackage{bm}
\usepackage{natbib}
\usepackage{braket}
\usepackage{float}
\usepackage{color}
\usepackage{blindtext}
\usepackage{comment}
\usepackage{xcolor}

\def\CRO{$\text{Ca}_2\text{RuO}_4$}

\begin{document}

\title{Time-hidden magnetic order in a multi-orbital Mott insulator}
	
\author{Xinwei Li}
\affiliation{Institute for Quantum Information and Matter, California Institute of Technology, Pasadena, CA 91125, USA}
\affiliation{Department of Physics, California Institute of Technology, Pasadena, CA 91125, USA}
\affiliation{Department of Physics, National University of Singapore, Singapore 117551, Singapore}
	
\author{Iliya Esin}
\affiliation{Institute for Quantum Information and Matter, California Institute of Technology, Pasadena, CA 91125, USA}
\affiliation{Department of Physics, California Institute of Technology, Pasadena, CA 91125, USA}
	
\author{Youngjoon Han}
\affiliation{Institute for Quantum Information and Matter, California Institute of Technology, Pasadena, CA 91125, USA}
\affiliation{Department of Physics, California Institute of Technology, Pasadena, CA 91125, USA}
	
\author{Yincheng Liu}
\affiliation{Institute for Quantum Information and Matter, California Institute of Technology, Pasadena, CA 91125, USA}
\affiliation{Department of Physics, California Institute of Technology, Pasadena, CA 91125, USA}
	
\author{Hengdi Zhao}
\affiliation{Department of Physics, University of Colorado, Boulder, CO 80309, USA}
	
\author{Honglie Ning}
\affiliation{Institute for Quantum Information and Matter, California Institute of Technology, Pasadena, CA 91125, USA}
\affiliation{Department of Physics, California Institute of Technology, Pasadena, CA 91125, USA}
	
\author{Cora Barrett}
\affiliation{Department of Physics, Wellesley College, Wellesley, MA 02481, USA}
	
\author{Jun-Yi Shan}
\affiliation{Institute for Quantum Information and Matter, California Institute of Technology, Pasadena, CA 91125, USA}
\affiliation{Department of Physics, California Institute of Technology, Pasadena, CA 91125, USA}
	
\author{Kyle Seyler}
\affiliation{Institute for Quantum Information and Matter, California Institute of Technology, Pasadena, CA 91125, USA}
\affiliation{Department of Physics, California Institute of Technology, Pasadena, CA 91125, USA}
	
\author{Gang Cao}
\affiliation{Department of Physics, University of Colorado, Boulder, CO 80309, USA}
	
\author{Gil Refael}
\affiliation{Institute for Quantum Information and Matter, California Institute of Technology, Pasadena, CA 91125, USA}
\affiliation{Department of Physics, California Institute of Technology, Pasadena, CA 91125, USA}
	
\author{David Hsieh}
\email[email: ]{dhsieh@caltech.edu}
\affiliation{Institute for Quantum Information and Matter, California Institute of Technology, Pasadena, CA 91125, USA}
\affiliation{Department of Physics, California Institute of Technology, Pasadena, CA 91125, USA}

\maketitle

\newpage

\newpage

\textbf{Photo-excited quantum materials can be driven into thermally inaccessible metastable states that exhibit structural, charge, spin, topological and superconducting orders. Metastable states typically emerge on timescales set by the intrinsic electronic and phononic energy scales, ranging from femtoseconds to picoseconds, and can persist for weeks. Therefore, studies have primarily focused on ultrafast or quasi-static limits, leaving an intermediate time window less explored. Here, we reveal a metastable glide symmetry-broken state in photo-doped \CRO\ using time-resolved optical second-harmonic generation and birefringence measurements. We find that the metastable state appears long after intra-layer antiferromagnetic order has melted and photo-carriers have recombined. Its properties are distinct from all known states in the equilibrium phase diagram and are consistent with intra-layer ferromagnetic order. Furthermore, model Hamiltonian calculations reveal that a non-thermal trajectory to this state can be accessed via photo-doping. Our results expand the search space for out-of-equilibrium electronic matter to metastable states emerging at intermediate timescales.}  

The multiband Mott insulator \CRO\ is an orthorhombic crystal built from two-dimensional (2D) planes of corner-sharing RuO$_6$ octahedra. The Ru$^{4+}$ ions (4d$^4$) are nominally 2/3-filled, with four electrons occupying the three $t_{2g}$ orbitals ($d_{xy}$, $d_{yz}$, $d_{xz}$). Strong coupling between charge, lattice, spin, and orbital degrees of freedom produces a cascade of phase transitions as a function of temperature ($T$): a metal-insulator transition at $T_\text{MIT}=357$~K, an orbital ordering transition at $T_\text{OO}=260$~K, and a Néel transition at $T_\text{N}=110$~K to a G-type antiferromagnetic (AFM) state. The structural driving force for these transitions is the compressive-type tetragonal distortion that flattens the RuO$_6$ octahedra, creating an energy offset ($\Delta > 0$) between the $d_{xy}$ orbital and the $d_{yz/xz}$ doublet (Fig.\,1a inset). Upon cooling across $T_\text{OO}$, there is a steep increase in $\Delta$ \cite{Braden1998,Zegkinoglou2005}, indicating its intimate connection to orbital order. Large $\Delta$ favors double occupancy of the $d_{xy}$ orbital and leaves the $d_{yz/xz}$ doublet half-filled, thereby promoting a Mott insulating G-type AFM ground state. Accounting for spin-orbit-coupling, the ground state can also be understood as an excitonic magnet arising from exchange-interaction-induced condensation of triplons \cite{Khaliullin2013,Souliou2017,Jain2017}. 

Previous idealized model Hamiltonian calculations for \CRO\ showed that when $\Delta$ is tuned below a critical value ($\Delta_c$), the energetically favored magnetic configuration in each 2D plane switches from being AFM with moments along the $b$-axis to ferromagnetic (FM) with moments along the $c$-axis \cite{Nomura2000,Fang2001,Meetei2015,Mohapatra2020} (Fig.\,1a). If anti-parallel stacking of FM planes is assumed, this corresponds to a transition from G-type AFM to A-type AFM ordering. To emphasize the change in ordering within a 2D plane, we hereafter refer to this transition as going from intra-layer AFM to intra-layer FM. The intra-layer FM state represents a local minimum in the potential energy landscape parameterized by $\Delta$ and the magnetic order parameter (OP) (Fig.\,1a), in which the system can potentially be trapped metastably. We propose that impulsive laser excitation can launch the system along a non-thermal trajectory from the intra-layer AFM state to the intra-layer FM state in the following way. In a magnetic Mott insulator, the sudden injection of photo-carriers can simultaneously decrease $\Delta$ by reducing the charge gap (see \cite{SM} Section~S2 for detailed calculations) and quench the magnetic OP \cite{Torre2022,Dean2016}. Beyond a critical excitation density where $\Delta$ transiently drops below $\Delta_c$, the system can transit through the intra-layer FM state before relaxing back to the intra-layer AFM state.

\begin{figure}[t!]
	\centering
	\includegraphics[width=0.95\linewidth]{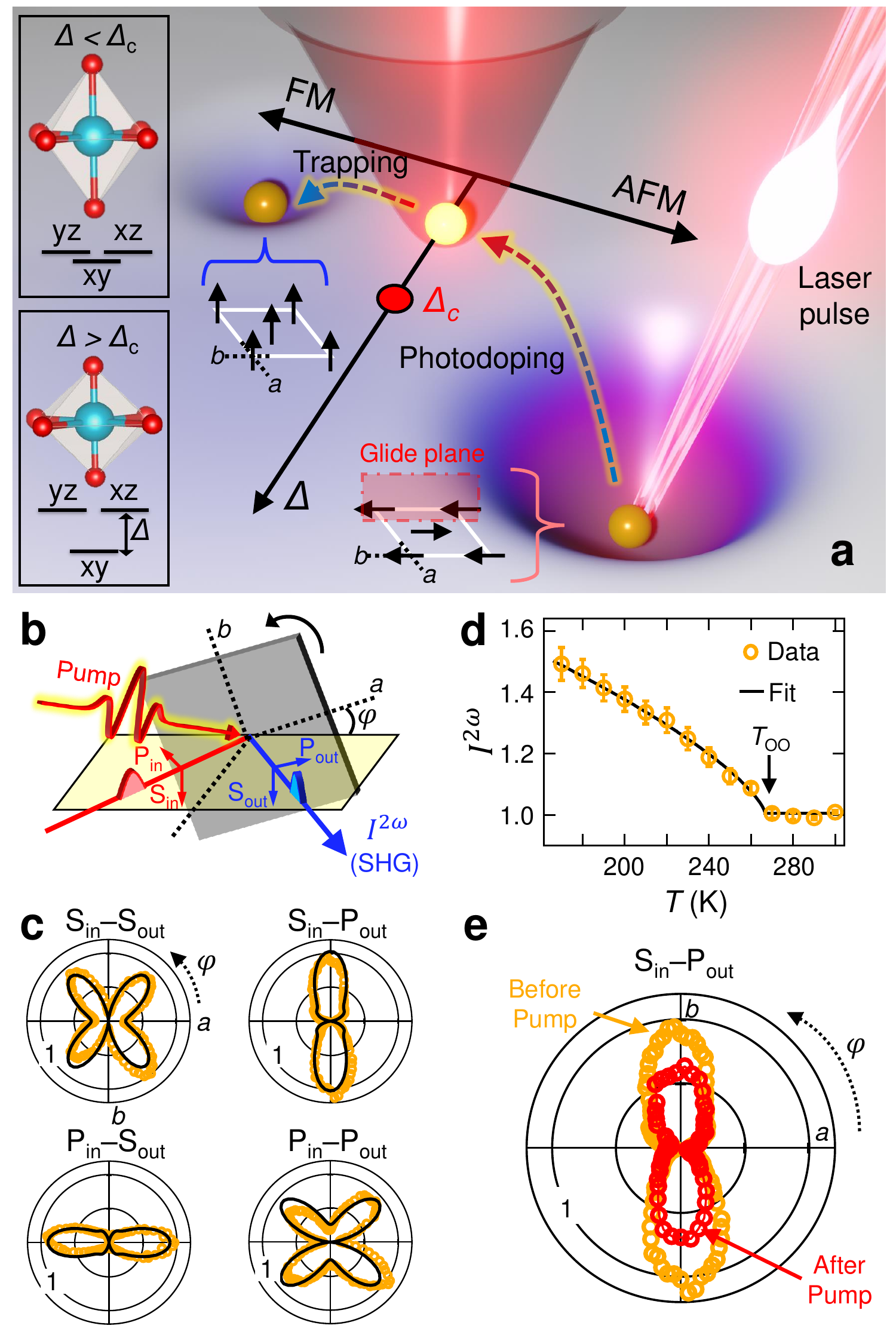}
	\caption{\small \textbf{Accessing the metastable magnetic state via photo-doping in \CRO.} \textbf{a},~Proposed trajectory (dashed arrows) in the potential energy landscape ($\Delta$ vs.~OP) from the intra-layer AFM to hidden intra-layer FM state launched by impulsive photo-doping. The moment configurations (small canting angles not depicted) in both states are illustrated, with the $bc$ glide plane marked. The red paraboloid shows the potential energy landscape in the transient paramagnetic state. Insets show schematic RuO$_6$ octahedral distortions and corresponding $d$-orbital energies for $\Delta>\Delta_c$ and $\Delta<\Delta_c$. \textbf{b},~Experimental configuration for time-resolved RA-SHG. P(S)-polarization is parallel (perpendicular) to the scattering plane. \textbf{c},~Polarization-resolved RA-SHG patterns at $T$=80~K in equilibrium. Black lines are symmetry-based fits to an orthorhombic point group (\cite{SM} Section~S3). \textbf{d},~$\varphi$-integrated SHG intensity of S$_\text{in}$--P$_\text{out}$ data versus $T$. Black line is a fit using the order parameter temperature scaling law $I^{2\omega}-1\propto|1-T/T_\text{OO}|^\beta$, where $\beta$ is the critical exponent. \textbf{e},~$I^{2\omega}(\varphi)$ plots for S$_\text{in}$--P$_\text{out}$ geometry before (orange) and 0.1~ps after (red) impulsive photo-doping at 80~K. Fluence: $n_\text{dh}=1.9\times10^{-2}$/Ru. Error bars are defined as $+/-$SD from the mean values.
	}
	\label{Fig1}
\end{figure}   

To experimentally verify that ultrafast photo-doping reduces $\Delta$, we performed rotational anisotropy (RA) second-harmonic generation (SHG) experiments \cite{Harter2015}. Obliquely incident probe light is focused onto a $c$-cut \CRO\ single crystal, and the intensity of frequency-doubled reflected light $I^{2\omega}$ is measured as a function of the angle $\varphi$ through which the scattering plane is rotated about the $c$-axis (Fig.\,1b). Figure\,1c shows RA patterns taken in equilibrium under distinct input and output polarization combinations, which clearly respect the $ac$ and $bc$ glide planes of the orthorhombic lattice (\cite{SM} Section~S3). The SHG signal in the paramagnetic (PM) regime ($T > T_\text{N}$) originates from time-reversal-even ($i$-type) SHG processes. Upon cooling below $T_\text{OO}$, we observe an OP-like upturn in the $i$-type SHG intensity (Fig.\,1d) with no change in the symmetry of the RA patterns. This is consistent with previous neutron and x-ray diffraction measurements, which reported a sharp rise in $\Delta$ below $T_\text{OO}$ with no change in crystallographic symmetry \cite{Braden1998,Zegkinoglou2005}. Leveraging this positive correlation between $I^{2\omega}$ and $\Delta$, we performed time-resolved RA-SHG measurements with a pump photon energy ($\sim$ 0.9~eV) tuned near the Mott gap of \CRO\ to study the impact of photo-carriers on $\Delta$. Immediately after pumping with fluence $F$ = 2~mJ/cm$^2$ [corresponding to a doublon-holon pair density $n_\text{dh}=1.9\times10^{-2}$/Ru (\cite{SM} Section~S4)], the RA-SHG patterns are uniformly reduced in intensity (Fig.\,1e), confirming that photo-carriers indeed decrease $\Delta$. Subsequently, $\Delta$ recovers on the picosecond timescale (\cite{SM} Section~S3).

\begin{figure*}[t!]
	\centering
	\includegraphics[width=\linewidth]{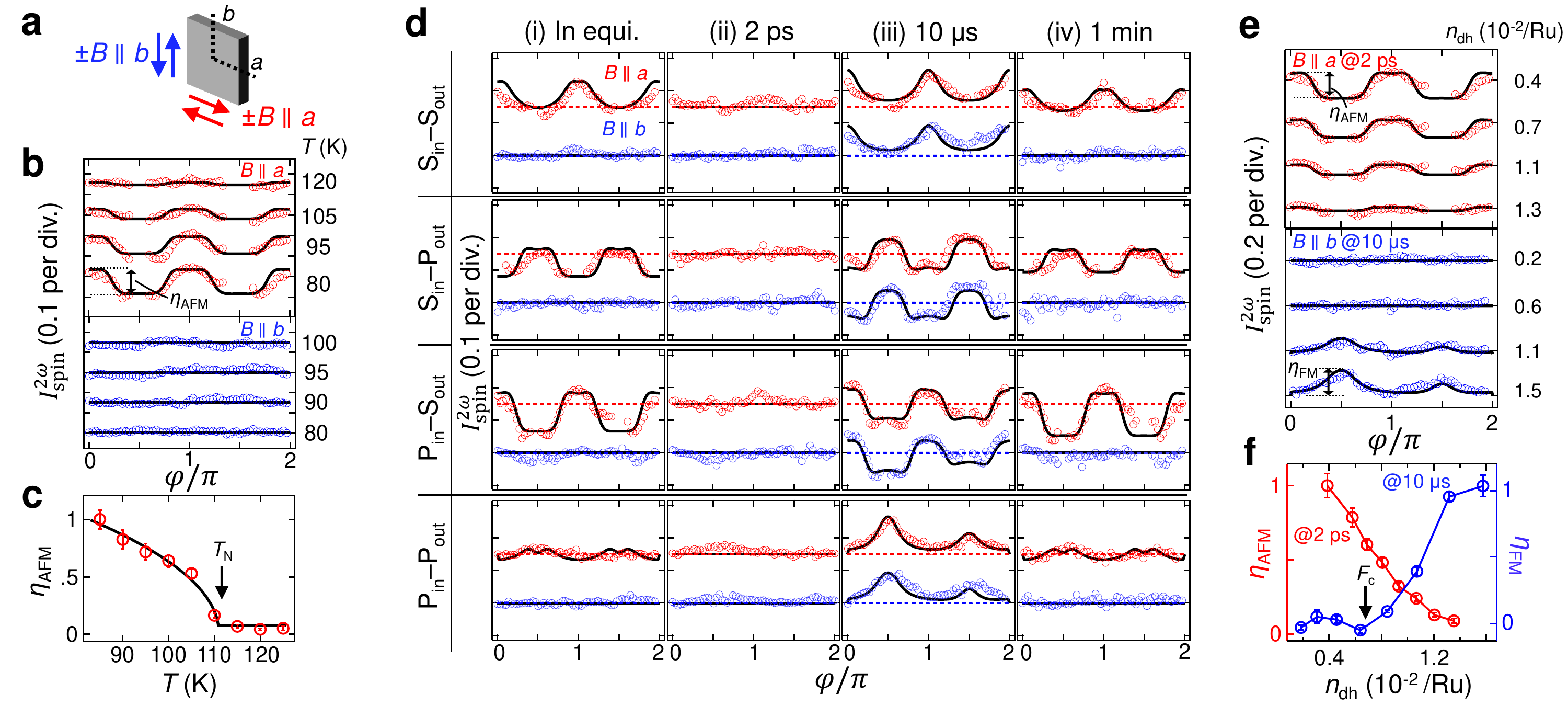}
	\caption{\small \textbf{Evidence for a metastable intra-layer ferromagnetic state from SHG.} \textbf{a},~Applied $B$-field directions to study $c$-type SHG response. \textbf{b},~$I^{2\omega}_\text{spin}(\varphi)$ for $B\parallel a$ (red curves) and $B\parallel b$ (blue curves) at various $T$ in P$_\text{in}$--S$_\text{out}$ geometry. Curves are vertically offset for clarity. The definition of $\eta_\text{AFM}$ is marked. \textbf{c},~$\eta_\text{AFM}$ versus $T$. Black line is a guide to the eye. \textbf{d},~Time evolution of $I^{2\omega}_\text{spin}(\varphi)$ for distinct polarization combinations and $B$ orientations. ``In equi." means sample is in equilibrium, when the pumping fluence is negligibly weak. ``1 min" snapshot obtained by abruptly shutting off an intense pump (that had induced metastable order at 10~$\upmu$s) and collecting data after 1 min wait time. \textbf{e},~$I^{2\omega}_\text{spin}(\varphi)$ for $B\parallel a$ (P$_\text{in}$--S$_\text{out}$) at 2 ps and for $B\parallel b$ (P$_\text{in}$--P$_\text{out}$) at 10~$\upmu$s at various pump fluences. Curves are vertically offset for clarity. The definition of $\eta_\text{FM}$ is marked. \textbf{f},~$\eta_\text{AFM}$ at 2~ps and $\eta_\text{FM}$ at 10~$\upmu$s versus $F$. For \textbf{b}, \textbf{d}, and \textbf{e}, black lines atop the $I^{2\omega}_\text{spin}(\varphi)$ data are symmetry-based fits to a surface electric-dipole SHG process with surface magnetic point group $m$, except for \textbf{d}(iii) and the lower panel of \textbf{e}, where we used surface magnetic point group $1$ (\cite{SM} Section~S6). Data points at those $\varphi$ angles where the total $I^{2\omega}$ is vanishingly small are discarded due to the ill-defined division process required to calculate $I^{2\omega}_\text{spin}$. $T=80$~K for \textbf{d}, \textbf{e}, and \textbf{f}. Error bars are defined as $+/-$SD from the mean values.
	}
	\label{Fig2}
\end{figure*}  

Below $T_\text{N}$, additional time-reversal-odd ($c$-type) SHG processes become allowed, which are sensitive to the symmetries of the magnetically ordered state. For an ideal collinear Néel state that is invariant under combined time-reversal and lattice translation operations, such contributions vanish by symmetry \cite{Torre2021}. However, in \CRO\, the moments are weakly canted away from the $b$-axis, producing a net in-plane magnetization ($\textit{\textbf{M}}$) along the $a$-axis \cite{Pincini2018}. Even if $\textit{\textbf{M}}$ reverses from layer-to-layer, a non-zero $c$-type electric-dipole surface SHG signal is in principle still allowed. This process is parameterized by a polar $c$-type susceptibility tensor $\chi^{(c)}_{ijk} \propto \textit{\textbf{M}}$ that governs the relationship between the induced polarization and incident electric field via $P_{\text{spin},i}^{2\omega}=\chi^{(c)}_{ijk}(\textit{\textbf{M}})E_jE_k$. To enhance sensitivity to this surface $c$-type response, we measured the difference between RA-SHG patterns acquired in opposite magnetic ($B$) fields (Fig.\,2a). In the limit where the surface $c$-type response is much smaller than the bulk $i$-type response, which is the typical case, the quantity $I_\text{spin}^{2\omega} =\frac{I^{2\omega}(+B)-I^{2\omega}(-B)}{I^{2\omega}(+B)+I^{2\omega}(-B)}\label{Ispindef}$ scales linearly with $\textit{\textbf{M}}$ (\cite{SM} Section~S5). 

Figure\,2b shows the temperature dependence of $I_\text{spin}^{2\omega}(\varphi)$ measured with the field ($|B|$ = 350 mT) aligned along either the $a$- or $b$-axis. Above $T_\text{N}$, $I_\text{spin}^{2\omega}(\varphi)$ is zero in both orientations, as expected for a paramagnet. Below $T_\text{N}$, a clear $\varphi$-dependent response emerges for $B\parallel a$, whereas none is detected for $B\parallel b$. This is expected because the moment canting angle is predominantly sensitive to fields applied perpendicular to the ordered moments. To further confirm the surface origin of the $c$-type SHG contribution, we note that the peak-to-peak amplitude of $I_\text{spin}^{2\omega}(\varphi)$ for $B\parallel a$ does not track the bulk magnetization \cite{Cao1997} but rather exhibits an OP-like onset at $T_\text{N}$ (Fig.\,2c), consistent with the behavior of $\textit{\textbf{M}}$ (\cite{SM} Section~S6). Moreover, $I_\text{spin}^{2\omega}(\varphi)$ data measured at $T < T_\text{N}$ under different polarization combinations [Fig.\,2d(i)] can all be simultaneously fit using a single susceptibility tensor respecting the surface magnetic point group $m$ of the intra-layer AFM structure (\cite{SM} Section~S6). Therefore, we take the amplitude of $I_\text{spin}^{2\omega}(\varphi)$ in the $B\parallel a$ channel as a measure of the intra-layer AFM OP ($\eta_\text{AFM}$).

The photo-doping induced dynamics of the magnetic OP can be tracked via the temporal evolution of $I_\text{spin}^{2\omega}(\varphi)$. Figure\,2d(ii) shows that following an intense pump pulse ($n_\text{dh}=1.6\times10^{-2}$/Ru, 100 fs pulse duration), the $I_\text{spin}^{2\omega}(\varphi)$ signal vanishes in all polarization channels by 2 ps, indicating a transient melting of the intra-layer AFM order. The system remains paramagnetic up to at least 100 ps (discussed later in Fig.\,4e), long after photo-dopants have recombined \cite{Li2022,Ning2023}. After blocking the pump beam with a mechanical shutter, the signals associated with intra-layer AFM order are fully recovered within 1 minute [Fig.\,2d(iv)]. Yet a surprising observation is made 10~$\upmu$s after the pump. This rarely studied intermediate timescale was accessed by delaying the probe pulse slightly before the pump pulse using a 100 kHz repetition rate laser \cite{Ravnik2021}. As shown in Fig.\,2d(iii), sizeable $I_\text{spin}^{2\omega}(\varphi)$ signals emerge in both the $B\parallel a$ and $B\parallel b$ orientations at this time. This cannot be reconciled with the equilibrium intra-layer AFM order [or a coexistence of different AFM domain types (\cite{SM} Section~S6)], but is consistent with $\textit{\textbf{M}}$ perpendicular to both the $a$- and $b$-axes. The $I_\text{spin}^{2\omega}(\varphi)$ curves at 10~$\upmu$s can only be fit by imposing a loss of the $bc$ glide plane (inset Fig.\,1a), which reduces the surface magnetic point group from $m$ to 1  (\cite{SM} Section~S6) - a symmetry change that is compatible with a transition from $b$-axis oriented intra-layer AFM to $c$-axis oriented intra-layer FM order. By spatially scanning the probe beam around the sample, we were able to identify domains with opposite sign of $\chi^{(c)}_{ijk}(\textit{\textbf{M}})$ (\cite{SM} Section~S11), further supporting an intra-layer FM state. We note that we did not detect any onset of magneto-optical Kerr rotation associated with the new order (\cite{SM} Section~S6), likely due to antiparallel stacking of the FM planes.

As shown in Figure\,2e, a finite $I_\text{spin}^{2\omega}(\varphi)$ signal in the $B\parallel b$ channel at 10~$\upmu$s only appears at large $F$ where the signal in the $B\parallel a$ channel at 2 ps has diminished, suggestive of competing OPs. Taking the peak-to-peak amplitude of $I_\text{spin}^{2\omega}(\varphi)$ for $B\parallel b$ as a measure of the intra-layer FM OP ($\eta_\text{FM}$), we overlay the $F$ dependence of $\eta_\text{AFM}$ and $\eta_\text{FM}$ (Fig.\,2f). We observe a clear inverse relationship, with $\eta_\text{FM}$ saturating after $\eta_\text{AFM}$ goes to zero. There also exists a critical fluence $F_\text{c}$ for $\eta_\text{FM}$ to onset, consistent with our conjecture in Fig.\,1a. We also performed $I_\text{spin}^{2\omega}(\varphi)$ measurements in the 0.1--100~ps time window but detected neither any finite $\eta_\text{FM}$ nor $\eta_\text{AFM}$ (see Fig.\,4e and later discussions), suggesting a slow appearance of the metastable order.

\begin{figure*}[t!]
	\centering
	\includegraphics[width=\linewidth]{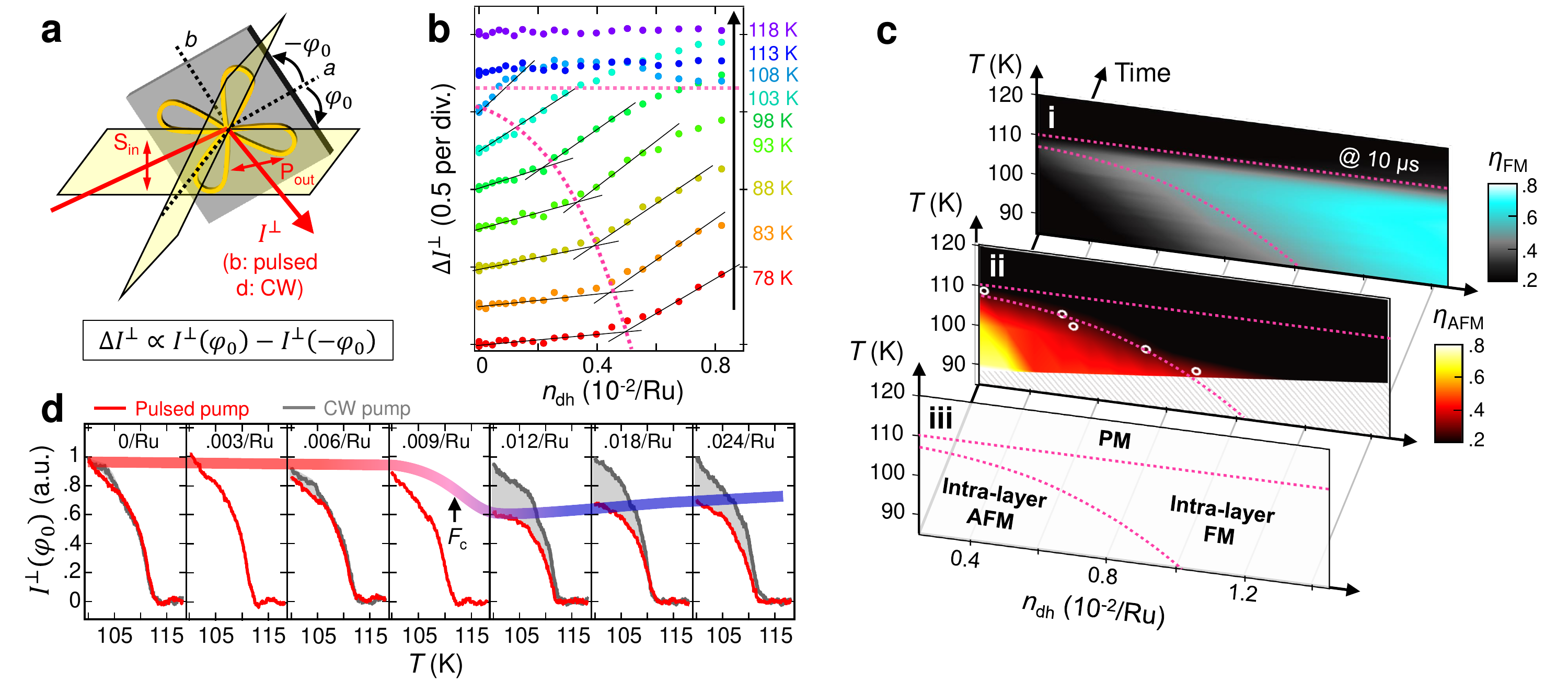}
	\caption{\small \textbf{Broken glide-plane symmetry in the metastable state probed by birefringence polarimetry.} \textbf{a},~Experimental configuration. Glide-symmetry-breaking is captured by subtracting the cross-polarized reflection intensity at two opposite scattering plane angles about the principal axes of \CRO. The yellow curve shows $I^{\perp}(\varphi)$ calculated for an orthorhombic crystal (\cite{SM} Section~S7). \textbf{b},~$F$ dependence of $\Delta I^{\perp}$ (measured at $\varphi=\pm\varphi_0=\pm20^{\circ}$) at various $T$'s. Black lines are guides to accentuate the change in slope. Upper (lower) dashed pink line charts the Néel temperature $T_\text{N}$ (temperature evolution of $F_c$). \textbf{c},~Out-of-equilibrium $T-F$ phase diagram mapped by (i) birefringence polarimetry at 10~$\upmu$s, (ii) SHG at 2 ps, and (iii) a superposition of both. Pink lines are the extracted phase boundaries from the $\eta_\text{AFM}$ map, which best matches the experimental data points (white circles, located at coordinates where $\eta_\text{AFM}$ decreases to 30\% of its 80-K value). The missing wedge-shaped region in (ii) is due to a temperature correction we applied to account for steady-state laser heating. \textbf{d},~$T$-dependence of $I^{\perp}(\varphi_0)$ measured by pulsed-pump CW-probe (red curves) or CW-pump CW-probe (grey curves) at various $F$'s. Within each panel, the pulsed pump and the CW pump have the same intensity. Line through all panels is a guide to the eye for locating $F_c$.
	}
	\label{Fig3}
\end{figure*}

To verify that the AFM-to-FM transition is a bulk effect, we searched for signatures of $bc$ glide symmetry breaking using bulk-sensitive birefringence polarimetry. For orthorhombic crystals, the cross-polarized reflection intensity is expected to vary as $I^\perp(\varphi)\propto(\chi_{aa}-\chi_{bb})^2\sin^2(2\varphi)$ (\cite{SM} Section~S7), where $\chi_{aa}$ and $\chi_{bb}$ are the diagonal elements of the linear electric susceptibility tensor and the scattering plane angle $\varphi$ is measured with respect to the $ac$-plane (Fig.\,3a). The difference in $I^\perp(\varphi)$ measured at two angles, $\varphi_0$ and $-\varphi_0$, symmetrical with respect to the $bc$-plane, $\Delta I^\perp \equiv I^\perp(+\varphi_0)-I^\perp(-\varphi_0)$, is zero (non-zero) if the $bc$ glide plane is present (absent). By adopting a modulation-based differential technique (\cite{SM} Section~S8), the instantaneous value of $\Delta I^\perp$ at 10~$\upmu$s as a function of both pump fluence and temperature was measured (Fig.\,3b). At $T$ = 78~K (below $T_\text{N}$), $\Delta I^\perp$ is near zero at low fluence and rises sharply once $F$ exceeds a critical value of approximately $n_\text{dh}=0.5\times10^{-2}$/Ru, closely following the SHG data (Fig.\,2f). Upon heating, $F_\text{c}$ progressively decreases and approaches zero near $T_\text{N}$. Above $T_\text{N}$, $\Delta I^\perp$ stays zero at all fluences, indicating that the metastable intra-layer FM state cannot be photo-induced from the PM state.

\begin{figure*}[thb]
	\centering
	\includegraphics[width=\linewidth]{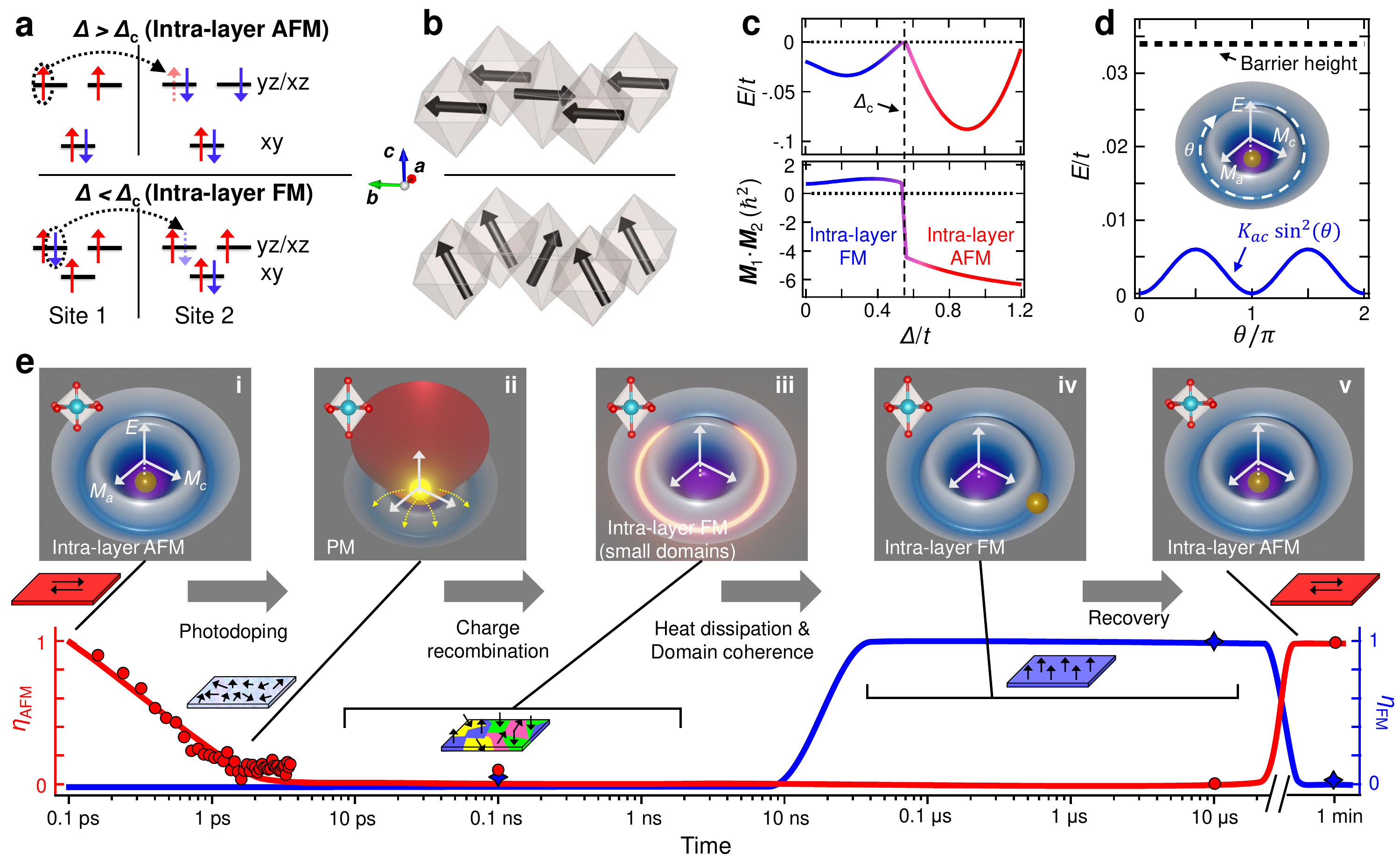}
	\caption{\small \textbf{Microscopic mechanism of the light-induced metastable state.} \textbf{a},~Two-site model of \CRO\ with $d^4$ filling. The lowest energy spin configurations (red and blue arrows) facilitating virtual hopping between sites (curved dashed arrow) are shown for $\Delta>\Delta_c$ (top panel) and $\Delta<\Delta_c$ (bottom panel). \textbf{b},~Schematics of the calculated intra-layer AFM and FM solutions. \textbf{c},~Calculated energy and inter-site magnetic correlation versus $\Delta$. \textbf{d},~Calculated magnetic anisotropy profile in the $ac$ plane in the {intra-layer FM} phase. The anisotropy energy $K_{ac}$ in the azimuthal ($\theta$) direction (blue curve) is much smaller than the energy barrier between the intra-layer AFM and FM states in the radial direction (dashed line). Inset: full energy landscape defined in the $M_a-M_c$ space. \textbf{e},~Upper panels: snapshots of the state population (yellow) and magnetic energy landscape at select times. (i) equilibrium intra-layer AFM state represented by an energy potential; (ii) immediately after photo-doping the system enters a PM state described by a paraboloidal potential; (iii) the potential is recovered but the system is trapped in the valley, possibly through the transient trapping mechanism proposed for systems with competing orders \cite{Sun2020}, forming small intra-layer FM domains; (iv) small intra-layer FM domains coalesce into larger domains; (v) relaxation back to intra-layer AFM state. The compressed (elongated) octahedron in the insets represents large (small) $\Delta$. Lower panel: temporal evolution of $\eta_\text{AFM}$ (red) and $\eta_\text{FM}$ (blue) inferred from experimental data points (circles and diamonds).
	}
	\label{Fig4}
\end{figure*}  

By collecting a detailed set of time-resolved $I_\text{spin}^{2\omega}(\varphi)$ data at 2 ps and $\Delta I^\perp$ data at 10~$\upmu$s as a function of fluence and temperature, we were able to construct a complete $T-F$ phase diagram for $\eta_\text{FM}$ and $\eta_\text{AFM}$ respectively [Figs\,3c(i) \& (ii)]. The phase boundaries (numerically extracted from the $\eta_\text{AFM}$ map) between the PM, intra-layer AFM, and intra-layer FM states are overlaid in Figure\,3c(iii). It is apparent that the emergence of intra-layer FM order in the intermediate time window is conditioned upon a photo-doping induced collapse of intra-layer AFM order on ultrashort timescales, supporting the proposed non-thermal trajectory in Fig.\,1a.   

Although it is challenging to map the complete temporal evolution of $\eta_\text{FM}$ due to the long timescales involved, we can ascertain its metastable nature by measuring $I^\perp(\varphi_0)$ with a continuous wave (CW) probe while irradiating the crystal with a train of pump pulses. With the pump beam off, the temperature dependence of $I^\perp(\varphi_0)$ exhibits a clear OP-like onset below $T_\text{N}$, in accordance with a coupling of intra-layer AFM order to birefringence (Fig.\,3d). Upon turning on the pump beam, there is little change in the temperature dependence of $I^\perp(\varphi_0)$ up to a pump fluence near $n_\text{dh} = 0.9\times10^{-2}$/Ru, above which the value of $I^\perp(\varphi_0)$ for $T < T_\text{N}$ suddenly drops. The fact that $F_\text{c}$ appears in a time-integrated (CW probe) measurement indicates that the transient intra-layer FM phase lasts for an appreciable fraction of the pump pulse train period (10 $\upmu$s). To rule out cumulative heating as a cause for the drop in $I^\perp(\varphi_0)$, we repeated the measurements using a CW pump (grey lines in Fig.\,3d) delivering the same average power as the pulse train. No changes in the $I^\perp(\varphi_0)$ versus temperature curves were observed in this case.

A microscopic understanding of our observations comes from multi-orbital Mott physics (Fig.\,4a). In the $\Delta = 0$ limit where the $d_{xz}$, $d_{yz}$ and $d_{xy}$ orbitals are degenerate, the Goodenough-Kanamori-Anderson rules stipulate that away from half-filling, FM correlations are energetically favored due to Hund's coupling \cite{Georges2013,Khomskii2022}. In the large $\Delta$ limit where orbitals are polarized and the $d_{yz/xz}$ orbitals are half-filled, AFM order is favored because it facilitates inter-site hopping. Therefore, a transition from intra-layer AFM to intra-layer FM order is expected as a function of $\Delta$. To study this transition more quantitatively, we developed a two-site model of \CRO\ that includes charge hopping, electron-lattice coupling (parameterized by $\Delta$), Kanamori interactions, and spin-orbit coupling (\cite{SM} Section~S9), and solved it by exact diagonalization. Our calculations show that below a certain $\Delta_\text{c}$, the ground state switches from being AFM ordered along the $b$-axis to being predominantly FM ordered along the $c$-axis - consistent with prior work \cite{Mohapatra2020} - albeit with a residual staggered in-plane moment (Fig.\,4b). 

Figure\,4c shows the calculated ground state energy and inter-site magnetic correlation versus $\Delta$. On either side of $\Delta_\text{c}$, there is a local and global energy minimum  characterized by FM and AFM correlations respectively. A potential barrier separates the two minima, representing intermediate spin and orbital configurations that are energetically less favorable.  Our results also reveal a weak magnetic anisotropy within the $ac$-plane in the intra-layer FM state. The energy landscape in the $M_a$-$M_c$ plane features a large barrier between the origin (intra-layer AFM state) and the valley (intra-layer FM state), but a much weaker barrier [$<$ 1\% of the hopping amplitude (\cite{SM} Section~S9)] along the azimuthal direction circling the valley (Fig.\,4d). This implies that when the system is excited from the intra-layer AFM to intra-layer FM state, all states around a ring in $M_a$-$M_c$-space are initially populated, corresponding to a patchwork of small domains with varying magnetic orientation. Crucially, the FM domain reorientation timescale is set by the anisotropy and is slow. Therefore, the time needed for FM domains to grow sufficiently large, such that a net magnetization is detectable in a spatially averaged measurement, can far exceed the timescale for intra-layer AFM order to collapse. 

Collectively, our experimental and theoretical results reveal the following temporal evolution (Fig.\,4e): (i) $t<0$: The system starts in equilibrium with $\Delta > \Delta_c$ and spatially uniform intra-layer AFM order. (ii) $t\sim1$~ps: Photo-doping with $F > F_c$ transiently drives $\Delta$ below $\Delta_c$ and melts intra-layer AFM order, turning the energy landscape to a paraboloid. (iii) 1~ps $<t<100$~ps: As the charge and spin subsystems rapidly cool and restore the original potential, the system preferentially relaxes to the intra-layer FM state, possibly through the transient trapping mechanism proposed for systems with competing orders \cite{Sun2020}. At this time, the intra-layer FM state exists in the form of small randomly oriented domains and both $\eta_\text{AFM}$ and $\eta_\text{FM}$ measure zero (see data in Fig.\,4e). (iv) $t>100$~ps: The small FM domains gradually coalesce into larger domains that can be optically resolved. (v) $t>1$~min: The system eventually overcomes the intra-layer FM to intra-layer AFM energy barrier via tunneling, nucleation, or domain wall motion \cite{Sun2020}, melting the metastable state and recovering intra-layer AFM order with the equilibrium value of $\Delta$.

Interestingly, this ``time-hidden" metastable intra-layer FM state has not been realized in thermal equilibrium in \CRO\ via any tuning parameter including temperature \cite{Braden1998}, carrier doping \cite{Cao2000,Pincini2018}, isovalent substitution \cite{Nakatsuji2000}, pressure \cite{Nakamura2002a,Steffens2005}, strain \cite{Ricco2018}, and electrical current \cite{Nakamura2013,Zhao2019}. Although itinerant ferromagnetism was reported under pressure \cite{Nakamura2002a,Steffens2005} and epitaxial strain \cite{Miao}, our light-induced intra-layer FM state emerges long after photo-dopants have recombined on the few picosecond timescale \cite{Li2022,Ning2023}. Additional transient optical reflectivity measurements show that \CRO\ remains Mott insulating above $F_c$ (\cite{SM} Section~S10), consistent with a recent work reporting a fluence threshold for the photo-induced insulator-to-metal transition that far exceeds $F_c$ \cite{Verma2024}. 

The mechanism uncovered in our work can in principle be applied generally to impulsively excited systems with competing order parameters, where domain coarsening dynamics following a quench can span a wide range of timescales \cite{Vogelgesang2018}. This vastly extends the search space for metastable states in strongly correlated quantum materials \cite{Liu2021,Morrison2014,Ichikawa2011,Stojchevska2014,Kogar2019,Yoshikawa2021,Duan2021,Han2023,Li2019,Nova2019,Li2013,Zhang2016,Disa2023,Afanasiev2021,Kimel2009,Sie2019,Zhang2019,Ning2022,Fausti2011,Mitrano2016} to intermediate timescales that have been previously overlooked. By employing a larger suite of impulsive excitation schemes, ranging from resonant phonon or magnon pumping to off-resonant Floquet band engineering \cite{Shan2021,Zhou2023,Afanasiev2021a}, potentially in combination, it may be possible to precisely control the trajectory of a material through phase space as well as the timing of when metastable states appear and disappear.

\section{Acknowledgements}
We thank Eugene Demler, Andrew Millis, Giniyat Khaliullin, and Wenyuan Liu for useful discussions. D.H. acknowledges support for instrumentation and theory from the Institute for Quantum Information and Matter (IQIM), an NSF Physics Frontiers Center (PHY-2317110). Optical spectroscopy measurements were funded in part by the Gordon and Betty Moore Foundation through Grant GBMF11564 to Caltech to support the work of D.H. G.R. expresses gratitude for the support by the Simons Foundation, the ARO MURI Grant
No. W911NF-16-1- 0361 and the Institute of Quantum
Information and Matter (PHY-2317110). X.L. acknowledges support from the Caltech Postdoctoral Prize Fellowship and the Singapore National Research Foundation under award no. NRF-NRFF16-2024-0008. G.C. acknowledges the support by National Science Foundation via grant no. DMR 2204811. I.E. acknowledges support from the Simons Foundation and the IQIM.

\section{Author Contributions}
X.L. and D.H. conceived the experiment. X.L., Y.H., and C.B. performed RA-SHG and BFISH experiments. X.L. and Y.L. performed birefringence polarimetry experiments. H.N. contributed to the understanding of the lattice SHG data. J.S. and K.S. developed the RA-SHG setup and adapted the setup for magnetic SHG measurements. K.S. made the magnetic rotator. I.E., G.R., and X.L. developed the theoretical interpretation. H.Z. and G.C. synthesized and characterized the samples.  D.H. supervised the project. X.L. and D.H. wrote the manuscript with input from all authors.
	
\section{Competing Interests}
The authors declare no competing interests.

\section{Materials and Methods}
\subsection{Material growth}

Single crystals of \CRO\ were synthesized using a NEC optical floating zone furnace with control of growth environment \cite{Cao2000,Qi2012}. The lateral dimension of the crystal used in the optics measurement was roughly 0.5~mm by 1~mm. The crystal was freshly cleaved along the $c$-axis right before the measurements and immediately pumped down to pressures better than $7\times10^{-7}$ torr in a continuous flow optical cryostat. Major observations reported in this manuscript have been reproduced at three different locations within the crystal. The critical fluence for the metastable transition can be different from location to location by a factor of 200$\%$ (highest divided by lowest).

\subsection{Optical measurements}

\subsubsection{Time-resolved second-harmonic generation polarimetry}
The time-resolved second-harmonic generation (SHG) rotational anisotropy (RA) setup is shown in Fig.\,1b of the main text. Our laser source was a Ti:sapphire-based regenerative amplifier (100~kHz, 800~nm, 100~fs), whose output beam was split into two arms: one feeding an optical parametric amplifier (OPA) to generate the pump beam (1400~nm), the other serving as the SHG-RA probe beam. Various pump-probe time delays were achieved by a mechanical optical delay line, where the 10~$\upmu$s snapshot was obtained by having the probe pulse arrive slightly before the pump which, in a repetitive measurement, is equivalent to setting the time delay to the laser pulsing period \cite{Ravnik2021}. The probe beam (800~nm) was focused on the crystal with at an incidence angle of 10$^\circ$ and a fluence of 0.8~mJ/cm$^{-1}$ (3 mW of probe power, 60~$\upmu$m diameter), and the intensity of the reflected frequency-doubled radiation (400~nm) was measured as a function of angle $\varphi$ through which the scattering plane is rotated about the surface normal. Various incident and output polarization combinations were used to sample different elements of the nonlinear susceptibility tensor. 

To increase the sensitivity of our apparatus to the $c$-type SHG response, magnetic-field-induced SHG (BFISH) measurements were performed by carrying out time-resolved SHG-RA measurements under in-plane magnetic ($B$) fields. The fields were applied by SmCo permanent magnets attached to a home-built rotator apparatus \cite{Seyler2020} placed adjacent to the sample on the cold finger of the cryostat. The field strength was fixed to 350~mT at the sample position and can be oriented $in$ $situ$ along any arbitrary in-plane direction. 

\subsubsection{Optical birefringence polarimetry}

The geometry of the optical birefringence polarimetry measurements is shown in Figure\,3a of the main text. The incident light is S-polarized and the intensity of P-polarized reflected light, $I^\perp(\varphi)$, is measured. For orthorhombic crystals, 
\begin{equation}
	I^\perp(\varphi)\propto(\chi_a-\chi_b)^2[\sin(2\varphi)]^2,
	\label{Iperp}
\end{equation}
where $\varphi$ is the scattering plane angle. Thus $I^\perp(\varphi)$ captures the difference in susceptibility along the two crystal axes, $\chi_a-\chi_b$, which is the birefringence. All measurements described below were repeated on a few spatial locations of the crystal to ensure the reported observations represent the behavior of the entire sample.

The modulation-based differential birefringence measurements (main text Fig.\,3) were carried out using 1300~nm pump light and 800~nm probe light. Birefringence measurements were carried out at two scattering plane angles by collecting $I^\perp(\varphi_0)$ and $I^\perp(-\varphi_0)$ ($\varphi_0=20^\circ$ defined from the crystal $a$ axis). $I^\perp(\varphi_0)$ and $I^\perp(-\varphi_0)$ are identical if there is a vertical glide plane (see Eq.\,\ref{Iperp}), but not if the glide plane is absent and the crystal symmetry is lowered from orthorhombic to monoclinic. As an optical chopper modulated the pump intensity, the pump-induced change in $I^\perp$, $dI^\perp$, was measured through lock-in detection. The reason why $dI^\perp$ measurements give symmetry information at 10~$\upmu$s is discussed in Supplemental Information.  

In time-averaged birefringence measurements (Fig.\,3d of main text), we replaced the pulsed probe beam by a continuous-wave 532~nm probe beam. The goal was to capture both the G-type antiferromagnetism (intra-layer AFM) onset and the light-induced glide-symmetry-breaking order, so temperature scanning at one of the scattering plane angles, $I^\perp(\varphi_0)$, was the priority. Two types of pumps were used: (1) pulses from the Ti-sapphire amplifier (100~kHz, 800~nm, 100~fs), and (2) a continuous-wave 785~nm laser beam. The continuous-wave pump provides similar focusing and local static heating conditions as the pulsed pump, but without impulsive charge carrier excitation.

\section{Data Availability}
The data that supports the plots within this paper has been provided. Data supporting other findings of this study are available from the corresponding author upon reasonable request.

\section{References in Methods}

	\newpage
\onecolumngrid	

\begin{center}
	\textbf{\large Supplementary Information for \\``Time-hidden magnetic order in a multi-orbital Mott insulator"}
\end{center}

\section{S1.~Materials and methods}
\subsection{A.~Material growth}

Single crystals of \CRO\ were synthesized using a NEC optical floating zone furnace with control of growth environment \cite{Cao2000,Qi2012}. The lateral dimension of the crystal used in the optics measurement was roughly 0.5~mm by 1~mm. The crystal was freshly cleaved along the $c$-axis right before the measurements and immediately pumped down to pressures better than $7\times10^{-7}$ torr in a continuous flow optical cryostat. Major observations reported in this manuscript have been reproduced at three different locations within the crystal. The critical fluence for the metastable transition can be different from location to location by a factor of 200$\%$ (highest divided by lowest).

\subsection{B.~Optical measurements}

\subsubsection{Time-resolved second-harmonic generation polarimetry}
The time-resolved second-harmonic generation (SHG) rotational anisotropy (RA) setup is shown in Fig.\,1b of the main text. Our laser source was a Ti:sapphire-based regenerative amplifier (100~kHz, 800~nm, 100~fs), whose output beam was split into two arms: one feeding an optical parametric amplifier (OPA) to generate the pump beam (1400~nm), the other serving as the SHG-RA probe beam. Various pump-probe time delays were achieved by a mechanical optical delay line, where the 10~$\upmu$s snapshot was obtained by having the probe pulse arrive slightly before the pump which, in a repetitive measurement, is equivalent to setting the time delay to the laser pulsing period \cite{Ravnik2021}. The probe beam (800~nm) was focused on the crystal with at an incidence angle of 10$^\circ$ and a fluence of 0.8~mJ/cm$^{-1}$ (3 mW of probe power, 60~$\upmu$m diameter), and the intensity of the reflected frequency-doubled radiation (400~nm) was measured as a function of angle $\varphi$ through which the scattering plane is rotated about the surface normal. Various incident and output polarization combinations were used to sample different elements of the nonlinear susceptibility tensor. 

To increase the sensitivity of our apparatus to the $c$-type SHG response, magnetic-field-induced SHG (BFISH) measurements were performed by carrying out time-resolved SHG-RA measurements under in-plane magnetic ($B$) fields. The fields were applied by SmCo permanent magnets attached to a home-built rotator apparatus \cite{Seyler2020} placed adjacent to the sample on the cold finger of the cryostat. The field strength was fixed to 350~mT at the sample position and can be oriented $in$ $situ$ along any arbitrary in-plane direction. 

\subsubsection{Optical birefringence polarimetry}

The geometry of the optical birefringence polarimetry measurements is shown in Figure\,3a of the main text. The incident light is S-polarized and the intensity of P-polarized reflected light, $I^\perp(\varphi)$, is measured. For orthorhombic crystals, 
\begin{equation}
	I^\perp(\varphi)\propto(\chi_a-\chi_b)^2[\sin(2\varphi)]^2,
	\label{Iperp}
\end{equation}
where $\varphi$ is the scattering plane angle. Thus $I^\perp(\varphi)$ captures the difference in susceptibility along the two crystal axes, $\chi_a-\chi_b$, which is the birefringence. All measurements described below were repeated on a few spatial locations of the crystal to ensure the reported observations represent the behavior of the entire sample.

The modulation-based differential birefringence measurements (main text Fig.\,3) were carried out using 1300~nm pump light and 800~nm probe light. Birefringence measurements were carried out at two scattering plane angles by collecting $I^\perp(\varphi_0)$ and $I^\perp(-\varphi_0)$ ($\varphi_0=20^\circ$ defined from the crystal $a$ axis). $I^\perp(\varphi_0)$ and $I^\perp(-\varphi_0)$ are identical if there is a vertical glide plane (see Eq.\,\ref{Iperp}), but not if the glide plane is absent and the crystal symmetry is lowered from orthorhombic to monoclinic. As an optical chopper modulated the pump intensity, the pump-induced change in $I^\perp$, $dI^\perp$, was measured through lock-in detection. The reason why $dI^\perp$ measurements give symmetry information at 10~$\upmu$s is discussed in Section~S8.  

In time-averaged birefringence measurements (Fig.\,3d of main text), we replaced the pulsed probe beam by a continuous-wave 532~nm probe beam. The goal was to capture both the G-type antiferromagnetism (intra-layer AFM) onset and the light-induced glide-symmetry-breaking order, so temperature scanning at one of the scattering plane angles, $I^\perp(\varphi_0)$, was the priority. Two types of pumps were used: (1) pulses from the Ti-sapphire amplifier (100~kHz, 800~nm, 100~fs), and (2) a continuous-wave 785~nm laser beam. The continuous-wave pump provides similar focusing and local static heating conditions as the pulsed pump, but without impulsive charge carrier excitation.

\section{S2.~Photocarrier-lattice coupling tunes orbital order} \label{SecMFmodel}

The electronic bands of \CRO\ near the Fermi level are dominated by the $t_{2g}$ orbitals of Ru$^{4+}$. The lowest-energy optical transition across the gap is assigned to be from the $d_{xy}$ to the $d_{yz/xz}$ orbital \cite{Fang2004}. According to prior photoemission spectroscopy measurements \cite{Sutter2017}, the experimental band structure can be best accounted for by a Mott gap imposed on $d_{yz/xz}$ bands due to onsite Coulomb repulsion energy $U$, a crystal-field induced downward shift of $d_{xy}$ bands by $\Delta$, and a splitting between the flat and fast-dispersing $d_{xy}$ bands by a few times the Hund's coupling energy $J_\text{H}$. Therefore, the optical gap size (band edge $d_{xy}\rightarrow d_{yz/xz}$ transition) is estimated to be $E_\text{gap}\sim U-2J_\text{H}+\Delta$. When carriers are created across the optical gap by ultrafast photodoping, the system tends to reduce $E_\text{gap}$ to minimize the potential energy. Since both $U$ and $J_\text{H}$ are atomic interaction parameters, reducing $\Delta$, a structural parameter, is the most viable way to reduce $E_\text{gap}$ in the nonequilibrium scenario, leading to transient weakening of the orbital order.

\begin{figure}[htb]
	\centering
	\includegraphics[width=\linewidth]{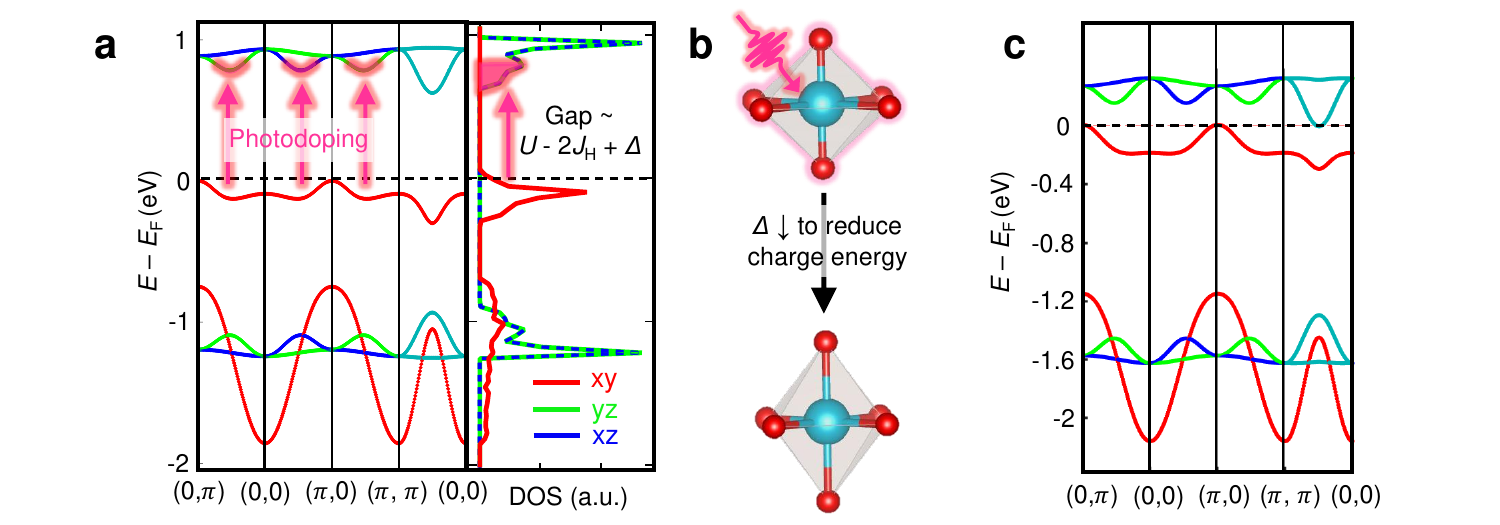}
	\caption{\textbf{a},~Calculated orbital-resolved band structure and density of states of \CRO\ by mean-field theory. \textbf{b},~Proposed mechanism of $\Delta$-tuning by photodoping. \textbf{c},~Band structure that is consistent with a nonequilibrium charge distribution. Details of the model and model parameters used to obtain this graph are described in the text.}
	\label{ChargelatticeMFT}
\end{figure}

We supplement the qualitative arguments above with mean-field theory calculations to further illustrate why photocarrier generation weakens the orbital order $\Delta$. Figure~\ref{ChargelatticeMFT} shows the calculated orbital-resolved electronic structure of \CRO\ by using a two-dimensional (2D) square-lattice three-band model, where we followed the treatment of Nomura \cite{Nomura2000}. The Hamiltonian consists of the one-body tight-binding part $H_0$ and electron-electron interaction part $H_\text{int}$:
\begin{align}
	H_0 & = \sum_{\textit{\textbf{k}},\alpha,\sigma}\epsilon_\alpha(\textit{\textbf{k}})c^\dagger_{\textit{\textbf{k}}\,\alpha\,\sigma}c_{\textit{\textbf{k}}\,\alpha\,\sigma}+\sum_{\textit{\textbf{k}},\sigma}\lambda(\textit{\textbf{k}})(c^\dagger_{\textit{\textbf{k}}\,yz\,\sigma}c_{\textit{\textbf{k}}\,xz\,\sigma}+c^\dagger_{\textit{\textbf{k}}\,xz\,\sigma}c_{\textit{\textbf{k}}\,yz\,\sigma}), \\
	H_\text{int} & = \frac{1}{2}U\sum_{i,\alpha,\sigma_1\neq\sigma_2}c^\dagger_{i\,\alpha\,\sigma_1}c^\dagger_{i\,\alpha\,\sigma_2}c_{i\,\alpha\,\sigma_2}c_{i\,\alpha\,\sigma_1}+
	\frac{1}{2}V\sum_{i,\alpha\neq\beta,\sigma_1,\sigma_2}c^\dagger_{i\,\alpha\,\sigma_1}c^\dagger_{i\,\beta\,\sigma_2}c_{i\,\beta\,\sigma_2}c_{i\,\alpha\,\sigma_1} \nonumber\\
	& +\frac{1}{2}J_\text{H}\sum_{i,\alpha\neq\beta,\sigma_1,\sigma_2}c^\dagger_{i\,\alpha\,\sigma_1}c^\dagger_{i\,\beta\,\sigma_2}c_{i\,\alpha\,\sigma_2}c_{i\,\beta\,\sigma_1},
	\label{meanfieldH}
\end{align}
where $\textit{\textbf{k}}=(k_x,k_y)$ is the crystal momentum, $i$, $j$ are site indices, $\alpha,\beta\in\{xy,yz,xz\}$ are orbital indices, and $\sigma,\sigma_1,\sigma_2\in\{+1,-1\}$ are spin indices. $c^\dagger$ and $n=c^\dagger c$ are the electron creation operator and number operator, respectively. The coefficients in $H_0$ are
\begin{align}
	\epsilon_{xy}(\textit{\textbf{k}}) & = -\frac{2}{3}\Delta+ 2t_1(\cos k_x+\cos k_y)+4t_2\cos k_x\cos k_y,\label{epsxy}\\
	\epsilon_{yz}(\textit{\textbf{k}}) & =\frac{1}{3}\Delta+2t_3\cos k_y+2t_4\cos k_x,\label{epsyz}\\
	\epsilon_{xz}(\textit{\textbf{k}}) & =\frac{1}{3}\Delta+2t_3\cos k_x+2t_4\cos k_y, \label{epsxz}\\
	\lambda(\textit{\textbf{k}}) & =4\lambda_0\sin k_x\sin k_y.
\end{align}
$t_1 - t_4$ and $\lambda_0$ are transfer integrals between nearest or next nearest neighbor Ru sites.

We used the Hartree-Fock approximation, where the operator products are replaced by $AB\rightarrow A\langle B\rangle+B\langle A\rangle-\langle A\rangle\langle B\rangle$, to treat $H_\text{int}$. We assume the following diagonal form of the mean field value:
\begin{equation}
	\langle c^\dagger_{i\,\alpha\,\sigma_1}c_{j\,\beta\,\sigma_2}\rangle=[n_\alpha+\frac{1}{2}\cos(\textit{\textbf{q}}\cdot\textit{\textbf{r}}_i)m_\alpha\sigma_1]\delta_{ij}\delta_{\alpha\beta}\delta_{\sigma_1\sigma_2},
\end{equation}
where $2n_\alpha$ is the number of electrons per site (or equivalently, per $\textit{\textbf{k}}$ point) occupying the orbital $\alpha$, $\textit{\textbf{q}}=(0,0)$ ($\textit{\textbf{q}}=(\pi,\pi)$) is the wavevector for ferromagnetic (antiferromagnetic) ordering, $m_\alpha$ is the magnetization carried by the orbital $\alpha$. The mean-field $H_\text{int}$ is then converted to $\textit{\textbf{k}}$-space through the Fourier transform
\begin{equation}
	c_i=\frac{1}{\sqrt{N}}\sum_{\textit{\textbf{k}}}^{\text{1\ BZ}}e^{i\textit{\textbf{k}}\cdot\textit{\textbf{r}}_i}c_\textit{\textbf{k}},
\end{equation}
where $N$ is the total number of atoms in real space (number of $\textit{\textbf{k}}$ points in reciprocal space). The total Hamiltonian can be written in the mean-field approximation as
\begin{align}
	H_\text{MF} & = \sum_{\textit{\textbf{k}},\alpha,\sigma}E_\alpha(\textit{\textbf{k}})c^\dagger_{\textit{\textbf{k}}\,\alpha\,\sigma}c_{\textit{\textbf{k}}\,\alpha\,\sigma}+\sum_{\textit{\textbf{k}},\sigma}\lambda(\textit{\textbf{k}})(c^\dagger_{\textit{\textbf{k}}\,yz\,\sigma}c_{\textit{\textbf{k}}\,xz\,\sigma}+c^\dagger_{\textit{\textbf{k}}\,xz\,\sigma}c_{\textit{\textbf{k}}\,yz\,\sigma})\nonumber\\
	& + \sum_{\textit{\textbf{k}},\alpha,\sigma}I_{\alpha\sigma}c^\dagger_{\textit{\textbf{k}}\,\alpha\,\sigma}c_{{\textit{\textbf{k}}+\textit{\textbf{q}}}\,\alpha\,\sigma} + C,
\end{align}
where
\begin{align}
	E_\alpha(\textit{\textbf{k}}) & = \epsilon_\alpha(\textit{\textbf{k}})+Un_\alpha+(2V-J_\text{H})\sum_{\beta\neq\alpha}n_\beta,\\
	I_{\alpha\sigma} & =-\frac{1}{2}(Um_\alpha+J_\text{H}\sum_{\beta\neq\alpha}m_\beta)\sigma,\\
	C & =-NU\sum_{\alpha}(n^2_\alpha-\frac{1}{4}m^2_\alpha)-2NV\sum_{\beta\neq\alpha}n_\alpha n_\beta \nonumber\\
	& +NJ_\text{H}\sum_{\beta\neq\alpha}(n_\alpha n_\beta+\frac{1}{4}m_\alpha m_\beta).
\end{align}

We used a 60$\times$60 $\textit{\textbf{k}}$ grid and therefore $N=3600$. We then assumed parameters applicable to \CRO: $t_1=-0.2$~eV, $t_2=0.375t_1$, $t_3=1.25t_1$, $t_4=0.125t_1$, $\lambda_0=0.2t_1$, $U=1.6$~eV, $J_\text{H}=0.4$~eV, $V=U-2J_\text{H}=0.8$~eV, $\Delta=0.72$~eV. Assuming the antiferromagnetic (AFM) solution, we solved eigen-energies and eigen-states for each $\textit{\textbf{k}}$ point, assumed a zero-temperature Fermi-Dirac distribution with a total filling of $N\sum_{\alpha}2n_\alpha=4N$, and solved $n_\alpha$ and $m_\alpha$ values in a self-consistent manner. The dimension of the Hilbert space for each $\textit{\textbf{k}}$ point is 12, considering the orbital and spin degrees of freedom, and the fact that $\textit{\textbf{k}}$ couples to $\textit{\textbf{k}}+\textit{\textbf{q}}$ in the AFM solution. Figure\,\ref{ChargelatticeMFT}a shows the calculated result of orbital-resolved band structure using the above conditions. The fact that the optical gap is contributed by the $d_{xy}\rightarrow d_{yz/xz}$ transition is apparent. Photodoping at our pump photon energy therefore depletes electrons in the $d_{xy}$ orbital and populates the $d_{yz/xz}$ orbital.

To demonstrate that generation of photo-carriers reduces $\Delta$, which is illustrated in Fig.\,\ref{ChargelatticeMFT}b, we modified the model so that $\Delta$ becomes a variable to be solved self-consistently. Following the treatment of Okamoto \cite{Okamoto2004}, we added an elastic lattice distortion energy term $H_\text{latt}=\frac{1}{2}K\Delta^2$ to the Hamiltonian ($K$ being the spring constant), and this term, together with the $\Delta$-dependent energy offset terms in Eq.\,\ref{epsxy}-Eq.\,\ref{epsxz}, constitutes an effective Jahn-Teller Hamiltonian:
\begin{equation}
	H_\text{JT}=\frac{1}{2}K\Delta^2+\frac{1}{3}(-2n_{xy}+n_{yz}+n_{xz})\Delta
\end{equation}
Through the saddle-point analysis of $H_\text{JT}$, $\Delta$ can be updated within each iteration so as to eventually give a self-consistent mean-field solution after many such iterations.

By picking $K=0.0926$~eV$^{-1}$, $\Delta$ converges to $\Delta=0.72$~eV,which gives the same band structure as in Fig.\,\ref{ChargelatticeMFT}a. Then we simulated the impact of photodoping by tuning the electron distribution function into a nonequilibrium profile. We achieved this by elevating a portion of the electrons (10\% of total population) from states immediately below the Fermi energy (valence band top) to states immediately above the Fermi energy (conduction band bottom), while fixing all other Hamiltonian parameters. Self-consistent solutions within this nonequilibrium charge distribution profile were then calculated. The solution converged to $\Delta=0.3$~eV for this case, representing a sizable decrease from the value ($\Delta=0.72$~eV) in equilibrium. Furthermore, the electronic band structure under the nonequilibrium charge distribution profile features a smaller optical gap, as shown in Fig.\,\ref{ChargelatticeMFT}c. All these results justify the arguments made earlier that illustrate how photocarriers impact $\Delta$ due to the need to minimize the nonequilibrium electronic energy. 

We next provide some insights into why a modified $\Delta$ is expected to change the experimental SHG intensity $I^{2\omega}$; the context is related to Figs.\,1d and e of the main text. $I^{2\omega}$ crucially depends on the nonlinear susceptibility tensor $\chi_{ijk}^{(2)}$, for which a quantum-mechanical calculation assuming a generic atomic model gives \cite{Boyd2008}
\begin{equation}
	\begin{split}
		\chi_{ijk}^{(2)} & \propto\sum_{m,n,v}{(\rho_{mm}^{(0)}-\rho_{vv}^{(0)})\frac{\mu_{mn}^i\mu_{nv}^j\mu_{vm}^k}{[\omega_{nm}-2\omega-i\gamma_{nm}][\omega_{vm}-\omega-i\gamma_{vm}]}}\\
		& -(\rho_{vv}^{(0)}-\rho_{nn}^{(0)})\frac{\mu_{mn}^i\mu_{vm}^j\mu_{nv}^k}{[\omega_{nm}-2\omega-i\gamma_{nm}][\omega_{nv}-\omega-i\gamma_{nv}]}
	\end{split}
\end{equation}
where $m$, $n$, $v$ are state labels, $\rho_{mm}$,  $\rho_{vv}$, and $\rho_{nn}$ are diagonal elements of density matrices specifying state populations, $\omega_{nm}$, $\omega_{vm}$, and $\omega_{nv}$ are atomic transition frequencies, $\gamma_{nm}$, $\gamma_{vm}$, and $\gamma_{nv}$ are transition linewidths, and $\mu_{nm}^i$, $\mu_{vm}^j$, and $\mu_{nv}^k$ are matrix elements of transition dipole moments. In continuous bands, the summation in the equation above would be replaced by an integral weighted by the joint density of states. With the increase of $\Delta$, the electronic band structure of \CRO\ changes; this would explicitly impact the values of $\rho$, $\omega$, $\gamma$, and $\mu$ in the integral, which in turn changes $I^{2\omega}$.

\section{S3.~Determining the crystallographic point group from SHG data}

\begin{figure}[htb]
	\centering
	\includegraphics[width=0.9\linewidth]{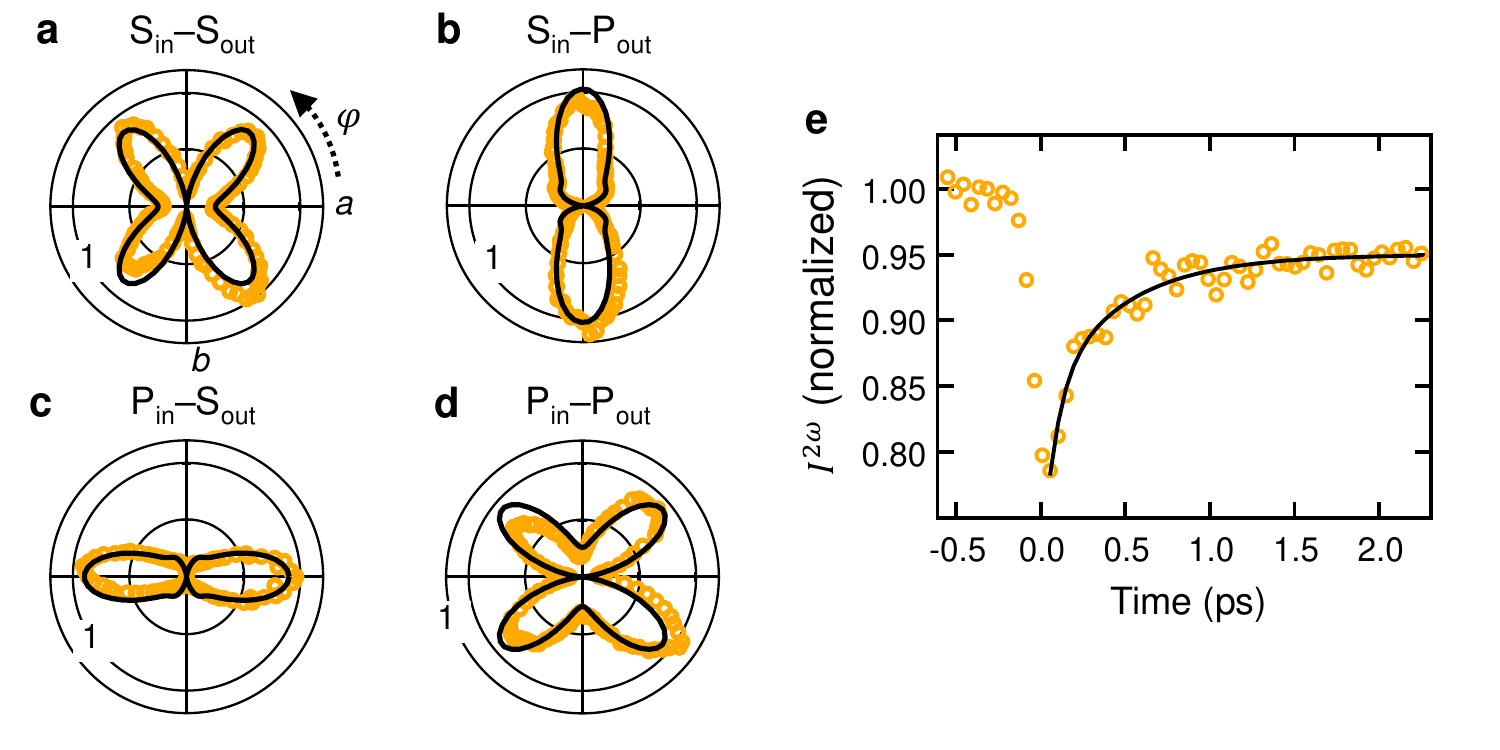}
	\caption{\textbf{a-d}, Polarization resolved lattice-induced RA-SHG patterns at 80~K in equilibrium. Black solid lines: model fit assuming the $mm2$ point group. \textbf{e}, Time-resolved SHG intensity of  P$_\text{in}$--P$_\text{out}$ channel pumped with $F=4$ mJ/cm$^2$, $n_\text{dh}=3.8\times10^{-2}$/Ru at 120~K. Black line: fit by exponential function with decay time of 0.5~ps.}
	\label{Latticepetalfit}
\end{figure}

We first describe the crystallographic lattice-induced ($i$-type) contribution to SHG from \CRO. Figure~1c of the main text (reproduced here as Fig.\,\ref{Latticepetalfit}a-d) shows the polarization resolved RA-SHG patterns at 80~K in equilibrium. The data were taken without an applied magnetic field. Since we did not observe any noticeable symmetry or intensity change of these patterns across $T_\text{N}$, we conclude that SHG is dominated by the lattice contribution even though the measurement temperature is below $T_\text{N}$.

The four input-output polarization channels in Fig.\,\ref{Latticepetalfit}a-d consistently show the orthorhombic symmetry of the crystal lattice - mirror axes of the patterns coincide with the $a$ and $b$ axes of the crystal. After photodoping, the time-resolved SHG response displayed in Fig.\,\ref{Latticepetalfit}e shows that the $i$-type SHG intensity recovers with an exponential time constant of around 1~ps.

To fit the RA-SHG data, we tried three possible crystallographic point groups belonging to the orthorhombic system: $mmm$, $mm$2, and $222$. For the centrosymmetric group $mmm$, we assumed four possible SHG processes:  
\\
\noindent (1) Bulk electric quadrupole (EQ) 
\begin{equation}
	P_{\text{latt},i}^{2\omega}\propto\chi_{ijkl}^\text{EQ}E_j^\omega\partial_kE_l^\omega,
	\label{bulkEQ}
\end{equation}
(2) Bulk magnetic dipole (MD) of MEE-type 
\begin{equation}
	M_{\text{latt},i}^{2\omega}\propto\chi_{ijk}^\text{MD1}E_j^\omega E_k^\omega,
	\label{bulkMD1}
\end{equation}
(3) Bulk magnetic dipole (MD) of PEH-type 
\begin{equation}
	P_{\text{latt},i}^{2\omega}\propto\chi_{ijk}^\text{MD2}E_j^\omega H_k^\omega,
	\label{bulkMD2}
\end{equation}
(4) Surface electric dipole (ED) 
\begin{equation}
	P_{\text{latt},i}^{2\omega}\propto\chi_{ijk}^\text{s}E_j^\omega E_k^\omega.
	\label{surfED}
\end{equation}
Here, $P^{2\omega}$ and $M^{2\omega}$ represent nonlinear polarization and magnetization, respectively. $\chi_{ijkl}^\text{EQ}$ is a polar $i$ tensor respecting the $mmm$ point group, $\chi_{ijk}^\text{MD1}$ and $\chi_{ijk}^\text{MD2}$ are both axial $i$ tensors respecting the $mmm$ point group, $\chi_{ijk}^\text{s}$ is a polar $i$ tensor respecting the surface group $mm2$ ($C_2\parallel c$). For both $mm2$ and $222$ groups, since they are noncentrosymmetric, we considered the bulk ED process
\begin{equation}
	P_{\text{latt},i}^{2\omega}\propto\chi_{ijk}^\text{ED}E_j^\omega E_k^\omega.
	\label{bulkED}
\end{equation}
where $\chi_{ijk}^\text{ED}$ is a polar $i$ tensor respecting the point group assumed.

We followed the approach outlined in Ref.\,\cite{Birss1966} to use the generators associated with the assumed point group to simplify the tensors. The elements within the simplified tensors were treated as variables to fit the experimental RA-SHG data. We found that the best fit is achieved by using the bulk ED term of the $mm2$ ($C_2\parallel b$) group, where the simplified tensor takes the form
\begin{equation}
	\chi_{ijk}^\text{ED}=
	{\scriptsize \begin{pmatrix}
			\begin{pmatrix}
				0\\\chi^\text{ED}_{xyx}\\0
			\end{pmatrix}
			\begin{pmatrix}
				\chi^\text{ED}_{xyx}\\0\\0
			\end{pmatrix}
			\begin{pmatrix}
				0\\0\\0
			\end{pmatrix}\\
			\begin{pmatrix}
				\chi^\text{ED}_{yxx}\\0\\0
			\end{pmatrix}
			\begin{pmatrix}
				0\\\chi^\text{ED}_{yyy}\\0
			\end{pmatrix}
			\begin{pmatrix}
				0\\0\\\chi^\text{ED}_{yzz}
			\end{pmatrix}\\
			\begin{pmatrix}
				0\\0\\0
			\end{pmatrix}
			\begin{pmatrix}
				0\\0\\\chi^\text{ED}_{zzy}
			\end{pmatrix}
			\begin{pmatrix}
				0\\\chi^\text{ED}_{zzy}\\0
			\end{pmatrix}
	\end{pmatrix}}
\end{equation}
\\
The calculated nonlinear polarization for the P$_\text{in}$--P$_\text{out}$, P$_\text{in}$--S$_\text{out}$, S$_\text{in}$--P$_\text{out}$, and S$_\text{in}$--S$_\text{out}$ channels, respectively, are
\begin{align}
	P^{2\omega}_{\text{latt,}pp} & \simeq \cos (\theta ) (-(\cos ^2(\theta ) ((2 \chi^\text{ED}_{xyx}+\chi^\text{ED}_{yxx}) \sin (\varphi ) \cos ^2(\varphi )+ \chi^\text{ED}_{yyy} \sin ^3(\varphi )))-\chi^\text{ED}_{yzz} \sin ^2(\theta ) \sin (\varphi )) \nonumber\\
	& +2 \chi^\text{ED}_{zzy} \sin ^2(\theta ) \cos (\theta ) \sin (\varphi ),\\
	P^{2\omega}_{\text{latt,}ps} & \simeq \cos ^2(\theta ) ((\chi^\text{ED}_{yyy}-2 \chi^\text{ED}_{xyx}) \sin ^2(\varphi ) \cos (\varphi )+\chi^\text{ED}_{yxx} \cos ^3(\varphi ))+\chi^\text{ED}_{yzz} \sin ^2(\theta ) \cos (\varphi ),\\
	P^{2\omega}_{\text{latt,}sp} & \simeq \cos (\theta ) (-(\chi^\text{ED}_{yyy}-2 \chi^\text{ED}_{xyx}) \sin (\varphi ) \cos ^2(\varphi )-\chi^\text{ED}_{yxx} \sin ^3(\varphi )),\\
	P^{2\omega}_{\text{latt,}ss} & \simeq (2 \chi^\text{ED}_{xyx}+\chi^\text{ED}_{yxx}) \sin ^2(\varphi ) \cos (\varphi )+\chi^\text{ED}_{yyy} \cos ^3(\varphi ),
\end{align}
where $\theta$ and $\varphi$ are the angle of incidence (fixed at 10$^\circ$ in measurements) and the scattering plane angle, respectively. Tensor elements $\chi^\text{ED}_{xyx}$, $\chi^\text{ED}_{yxx}$, $\chi^\text{ED}_{yyy}$, $\chi^\text{ED}_{yzz}$, and $\chi^\text{ED}_{zzy}$ are adjusted to obtain a simultaneous fit of $I^{2\omega}_{pp}\propto|P^{2\omega}_{\text{latt,}pp}|^2$, $I^{2\omega}_{ps}\propto|P^{2\omega}_{\text{latt,}ps}|^2$, $I^{2\omega}_{sp}\propto|P^{2\omega}_{\text{latt,}sp}|^2$, $I^{2\omega}_{ss}\propto|P^{2\omega}_{\text{latt,}ss}|^2$ to the experimental patterns; the result is shown in Fig.\,\ref{Latticepetalfit}a-d. The fitting works well across the orbital ordering temperature and up to the highest temperature accessible by our experiment. 

\begin{table}[htb]
	\begin{center}
		\begin{tabular}{|c|c|c|}
			\hline \textbf{point group} & \textbf{Type of radiation} & \textbf{Reason for failure of fitting}\\
			\hline
			$mm2$ ($C_2\parallel b$) & Bulk-ED & fits well\\
			\hline
			$mmm$ & EQ, MEE-MD, PEH-MD & predicts $I^{2\omega}_{ps}(\varphi=0)=I^{2\omega}_{ps}(\varphi=\pi/2)=0$\\
			\hline
			$mmm$ & Surface-ED & predicts $I^{2\omega}_{ss}=0$\\
			\hline
			$222$ & Bulk-ED & predicts $I^{2\omega}_{ss}=0$\\
			\hline
			$mm2$ ($C_2\parallel c$) & Bulk-ED & predicts $I^{2\omega}_{ss}=0$\\
			\hline
			$2/m$ ($C_2\parallel c$) & EQ, MEE-MD, PEH-MD & predicts no mirror axis in SHG patterns\\
			\hline
			$2/m$ ($C_2\parallel c$) & Surface-ED & predicts $I^{2\omega}_{ss}=0$\\
			\hline
			$2/m$ ($C_2\parallel b$) & EQ, MEE-MD, PEH-MD, Surface-ED & predicts no mirror axis in birefringence patterns\\
			\hline
		\end{tabular}
	\end{center}
	\caption{Fitting of lattice SHG patterns by distinct radiation processes of various crystallographic point groups. $mm2$ with $C_2\parallel b$ provides the unambiguous best fit. Reasons for the failures of fitting by other groups are listed.}
	\label{Tbllatticefit}
\end{table}

Table\,\ref{Tbllatticefit} summarizes the fits to all other crystallographic point groups (also including some tests with monoclinic point groups), which show significant disagreement with data. Considering the key points of discrepancy listed in the table, neither these processes themselves, nor coherent sums of different processes within any one of these point groups, can explain our data. It should be noted that for $mmm$, which is the most widely reported point group for \CRO, none of the four tested radiation processes are able to explain the data: the EQ, MEE-MD and PEH-MD processes all predict $I^{2\omega}_{ps}(\varphi=0)=I^{2\omega}_{ps}(\varphi=\pi/2)=0$, which contradicts Fig.\,\ref{Latticepetalfit}c. The Surface-ED radiation of $mmm$ originates from the $mm2$ group with $C_2\parallel c$. This would give $I^{2\omega}_{ss}=0$, which also contradicts the data. Our findings are consistent across multiple samples originating from different growth batches. 

We also attempted to use other index-two subgroups of $mmm$. For $222$ and $mm2$ ($C_2\parallel c$), the prediction from the Bulk-ED process is inconsistent with the $I^{2\omega}_{ss}$ data. For $2/m$, we considered two possible orientations: $C_2\parallel c$ and $C_2\parallel b$. Due to the monoclinic symmetry of the group, it predicts no common mirror axes in the four polarization channels of SHG, or in the linear birefringence. The fact that our data clearly show mirror planes rules out these point groups.

\section{S4.~Conversion between pump fluence and doublon-holon pair density}

Since photodoping is the main driving mechanism, a broad range of pump photon energies above the optical gap of \CRO\ can be applied to induce the metastable magnetic phase transition. This agrees with our experimental observation. Time-resolved SHG measurements were carried out with 1400~nm (0.89~eV) pumping. Time-resolved birefringence polarimetry measurements were carried out with 1300~nm (0.95~eV) pumping. Time-averaged birefringence measurements were carried out with 800~nm (1.55~eV) pumping. For proper comparison of fluence-dependent data and phase diagrams across these experimental runs, we converted fluence to doublon-holon (d-h) pair density.

The formula for calculating pump fluence (mJ/cm$^2$) is 
\begin{equation}
	F=P/(f_\text{rep}A)
\end{equation}
where $P$ (W) is the pump power, $f_\text{rep}$ (Hz) is the laser repetition rate, and $A$ (cm$^2$) is the area of the pumped laser spot. The d-h pair density is given by
\begin{equation}
	n_\text{dh}=\frac{F(1-R)V_\text{uc}}{N_\text{Ru}E_\text{ph}\delta}
\end{equation}
where $R$ and $\delta$ are, respectively, the reflectivity and penetration depth at the pump photon energy, which are determined by prior broadband optical conductivity measurements \cite{Li2022}. $V_\text{uc}$ is the unit cell volume, $N_\text{Ru}$ is the number of Ru sites per unit cell, and $E_\text{ph}$ is the pump photon energy.

\section{S5.~Magnetic-field-induced second-harmonic generation} \label{SecBFISH}

The onset of intra-layer AFM order in \CRO\ is not accompanied by any crystallographic symmetry breaking \cite{Braden1998}. Since the $i$-type crystallographic contribution dominates the SHG signal, no measurable change in the symmetry or intensity of the RA patterns was detected across $T_\text{N}$. To boost sensitivity to magnetic order-induced $c$-type SHG contributions, we used the magnetic-field-induced second-harmonic generation (BFISH) technique \cite{Fiebig2005}. The mathematical framework for this technique is described here. 

The canting-induced in-plane magnetization $\textit{\textbf{M}}$ turns on a $c$-type Surface-ED SHG process: $P_{\text{spin},i}^{2\omega}\propto\chi_{ijk}^\text{(c)}(\textit{\textbf{M}})E^{\omega}_jE^{\omega}_k$, where $\chi_{ijk}^\text{(c)}(\textit{\textbf{M}})$ is a third-rank polar $c$ tensor that is proportional to the magnetic order parameter (MOP). The reason why we focus on Surface-ED rather than Bulk-ED or Bulk-MD SHG is explained in the next Section. The spin-induced $c$-type polarization $P_{\text{spin},i}^{2\omega}$, which flips sign if $\textit{\textbf{M}}$ is reversed, can be expressed as follows:
\begin{align}
	P_{\text{spin},i}^{2\omega} & \propto\chi_{ijk}^\text{(c)}(\textit{\textbf{M}})E^{\omega}_jE^{\omega}_k\\
	& =\chi_{ijkl}^\text{(i)}M_lE^{\omega}_jE^{\omega}_k\label{BFISHM}\\
	& =\chi_{ijkl}^\text{(i)}\chi_{lp}^\text{m}B_pE^{\omega}_jE^{\omega}_k\\
	& =\chi_{ijkp}^\text{BFISH}B_pE^{\omega}_jE^{\omega}_k\\
	& =\chi_{ijkl}^\text{BFISH}B_lE^{\omega}_jE^{\omega}_k
	\label{BFISH1}
\end{align}
where both $\chi_{ijkl}$ and $\chi_{ijkl}^\text{BFISH}$ are fourth-rank axial $i$ tensors, $\chi_{lp}^\text{m}$ is a second-rank polar $i$ tensor, and $M_{l}$ and $B_{l/p}$ are magnetization and $B$-field components, respectively. In this derivation we used the expansion $\chi_{ijk}^\text{(c)}(\textit{\textbf{M}})=\chi_{ijkl}^\text{(i)}M_l$ and the $M$-$B$ relation $M_l=\chi_{lp}^\text{m}B_p$. The tensor $\chi^\text{BFISH}_{ijkp}=\chi_{ijkl}^\text{(i)}\chi_{lp}^\text{m}$, being the product of an axial $i$ tensor and a polar $i$ tensor, is an axial $i$ tensor. In additional to being proportional to the MOP, it bridges the BFISH process (Eq.\,\ref{BFISH1}) where two fundamental electric fields and one static $B$-field combine to give the spin-induced $c$-type polarization $P_{\text{spin},i}^{2\omega}$, which switches sign upon $B$-field reversal or $\textit{\textbf{M}}$ reversal: $P_\text{spin}^{2\omega}(B)=-P_\text{spin}^{2\omega}(-B)$.

The spin-induced $P_{\text{spin}}^{2\omega}$ polarization interferes with the lattice SHG polarization $P^{2\omega}_\text{latt}$ as given in Eq.\,\ref{bulkED}, giving the total polarization
\begin{equation}
	P^{2\omega}_\text{tot}(B)=P^{2\omega}_\text{latt}+P_{\text{spin}}^{2\omega}(B).
\end{equation}
Such interference enables us to gain direct information about $P_{\text{spin}}^{2\omega}(B)$, which is proportional to the MOP. We achieved this by successively measuring the SHG intensity in opposite fields ($B$ and $-B$). Since
\begin{align}
	I^{2\omega}(B) & \propto|P^{2\omega}_\text{tot}(B)|^2=|P^{2\omega}_\text{latt}|^2+|P_{\text{spin}}^{2\omega}(B)|^2+2P^{2\omega}_\text{latt}P_{\text{spin}}^{2\omega}(B) \label{IplusB}\\
	I^{2\omega}(-B) & \propto|P^{2\omega}_\text{tot}(-B)|^2=|P^{2\omega}_\text{latt}|^2+|P_{\text{spin}}^{2\omega}(-B)|^2+2P^{2\omega}_\text{latt}P_{\text{spin}}^{2\omega}(-B), \label{IminusB}
\end{align}
we can construct a quantity $I_\text{spin}^{2\omega}$
\begin{align}
	I_\text{spin}^{2\omega} & =\frac{I^{2\omega}(+B)-I^{2\omega}(-B)}{I^{2\omega}(+B)+I^{2\omega}(-B)} \simeq2P_{\text{spin}}^{2\omega}(B)/P^{2\omega}_\text{latt}
	\label{BFISHIspin}
\end{align}
that is proportional to $P_{\text{spin}}^{2\omega}(B)$ and therefore the nonzero components within the $\chi^\text{BFISH}_{ijkp}$ tensor. Further, due to the normalization factor $P^{2\omega}_\text{latt}$, $I_\text{spin}^{2\omega}$ is an intrinsic material property independent of probe laser intensity variations between different measurements. Note that the equation above holds when $|P_{\text{spin}}^{2\omega}(B)|\ll|P^{2\omega}_\text{latt}|$, which is true in our data. The derivation of Eq.\,\ref{BFISHIspin} assumed $P^{2\omega}_\text{latt}$ and $P_{\text{spin}}^{2\omega}(B)$ to take real values. Generalizing to cases where they are complex functions, we obtain the spin-sensitive component
\begin{equation}
	I_\text{spin}^{2\omega}\simeq2P_{\text{spin}}^{2\omega}(B)/P^{2\omega}_\text{latt}+2[P_{\text{spin}}^{2\omega}(B)]^*/(P^{2\omega}_\text{latt})^*,
	\label{cmplxBFISHIspin}
\end{equation}  
which still couples to linear order to components of the $\chi^\text{BFISH}_{ijkp}$ tensor, although coherent mixing of real and imaginary parts of tensor elements might occur. $I^{2\omega}_\text{spin}$ is therefore always a faithful reporter of the magnetic order parameter.

The rotational anisotropy of $I_\text{spin}^{2\omega}$, that is, $I_\text{spin}^{2\omega}(\varphi)$, is compared between experiment and model calculations to determine the tensor structure of $\chi^\text{BFISH}_{ijkl}$, which gives the magnetic point group symmetry of the spin order. This process is detailed in the next Section.

\section{S6.~Fits to the BFISH data}
\label{SecBFISH2}

First, we note that, by examining the magnetic unit cells of the two widely reported magnetic structures of \CRO, the magnetic point groups for the A-centered intra-layer AFM and B-centered intra-layer AFM structures are $mm2$ and $m'm2'$, respectively. This is assumed in all the symmetry analysis of tensors discussed below. Then, we address how we determine that it is a Surface-ED process rather than a bulk SHG process that underlies the spin-induced nonlinear polarization $P_{\text{spin}}^{2\omega}$. (1) The intra-layer AFM order parameter ($\eta_\text{AFM}$ plotted in Fig.\,2c of the main text), which is proportional to $P_{\text{spin}}^{2\omega}$ and probed by BFISH, shows an onset with monotonic temperature dependence below $T_\text{N}$. Given Eq.\,\ref{BFISHM}, if $P_{\text{spin}}^{2\omega}$ were probing bulk magnetization, the quantity would show a pronounced non-monotonic cusp following the temperature-dependent magnetization of \CRO\ in the low-$B$ regime (where $M$ is linearly proportional to $B$ and $B$ is far below the spin-flop field) \cite{Cao2000}. Instead, the observed behavior can only be explained if $P_{\text{spin}}^{2\omega}$ couples to surface magnetization. Although the magnetization between different RuO$_6$ layers mutually cancel, the magnetization in a single RuO$_6$ layer is finite, and scales with the magnitude of the AFM staggered moment. (2) By applying the generators of the bulk $mm2$ and $m'm2'$ point groups, $c$-type Bulk-ED radiation should yield $P_{\text{spin}}^{2\omega}(B)=0$ for the S$_\text{in}$--S$_\text{out}$ channel, which does not agree with our data (Fig.\,2d of main text). We further confirmed that $c$-type Bulk-MD radiation, which is reported for Sr$_2$IrO$_4$ in an external magnetic field \cite{Seyler2020}, cannot explain the same data set either (especially the $B\parallel b$ channel at high fluence) due to insufficient number of fitting parameters. Ruling out these candidates helps us rule out the occurrence of a B-centered intra-layer AFM structure of \CRO\ \cite{Braden1998}, for which $M$ is aligned parallel between layers, in both the equilibrium state and the high-fluence metastable state in our sample. 

\subsection{A.~BFISH from intra-layer AFM phase}
We consider the Surface-ED BFISH as described in Eq.\,\ref{BFISH1}:
\begin{align}
	P_{\text{spin},i}^{2\omega} & \propto\chi_{ijkl}^\text{BFISH}E^{\omega}_jE^{\omega}_kB_l.
	\label{BFISH2}
\end{align}
From our fits to the lattice RA-SHG data, we determined the structural point group to be $mm2$ ($C_2\parallel b$). This means that in the intra-layer AFM phase, the surface magnetic point group is $m$ ($\sigma_{bc}$), which contains only one glide plane (crystallographic $bc$ plane).

\begin{figure}[htb]
	\centering
	\includegraphics[width=\linewidth]{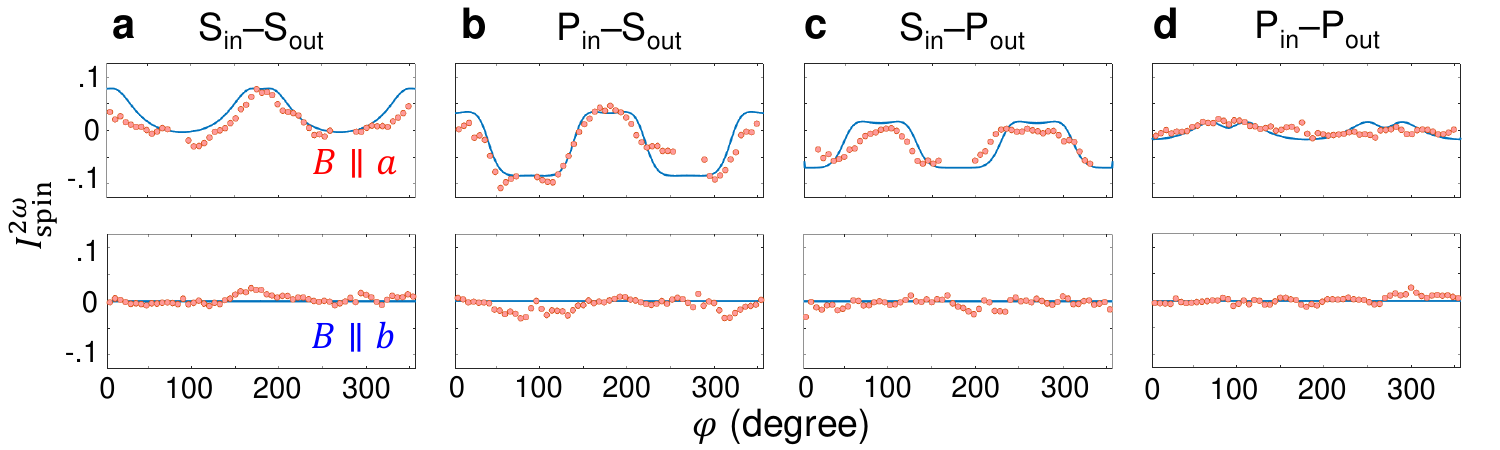}
	\caption{\textbf{Fits to BFISH data in the intra-layer AFM phase in equilibrium.} \textbf{a},~S$_\text{in}$--S$_\text{out}$, \textbf{b},~P$_\text{in}$--S$_\text{out}$, \textbf{c},~S$_\text{in}$--P$_\text{out}$, \textbf{d},~P$_\text{in}$--P$_\text{out}$. Top (lower) row panels all correspond to the $B\parallel a$ ($B\parallel b$) configuration. Red circles: experimental data. Blue lines: model calculation.}
	\label{BFISHfitGAFM}
\end{figure}

By applying the symmetry element of $\sigma_{bc}$ and the intrinsic symmetry $\chi_{ijkl}^\text{BFISH}=\chi_{ikjl}^\text{BFISH}$, we simplified the $\chi_{ijkl}^\text{BFISH}$ tensor as follows
\\
\begin{equation}
	\chi_{ijkl}^\text{BFISH}=
	{\scriptsize \begin{pmatrix}
			\begin{pmatrix}
				0 & \chi^\text{BFISH}_{xxxy} & \chi^\text{BFISH}_{xxxz}\\\chi^\text{BFISH}_{xyxx} & 0 & 0\\\chi^\text{BFISH}_{xzxx} & 0 & 0
			\end{pmatrix}
			\begin{pmatrix}
				\chi^\text{BFISH}_{xyxx} & 0 & 0\\0 & \chi^\text{BFISH}_{xyyy} & \chi^\text{BFISH}_{xyyz}\\0 & \chi^\text{BFISH}_{xzyy} & \chi^\text{BFISH}_{xzyz}
			\end{pmatrix}
			\begin{pmatrix}
				\chi^\text{BFISH}_{xzxx} & 0 & 0\\0 & \chi^\text{BFISH}_{xzyy} & \chi^\text{BFISH}_{xzyz}\\0 & \chi^\text{BFISH}_{xzzy} & \chi^\text{BFISH}_{xzzz}
			\end{pmatrix}\\
			\begin{pmatrix}
				\chi^\text{BFISH}_{yxxx} & 0 & 0\\0 & \chi^\text{BFISH}_{yyxy} & \chi^\text{BFISH}_{yyxz}\\0 & \chi^\text{BFISH}_{yzxy} & \chi^\text{BFISH}_{yzxz}
			\end{pmatrix}
			\begin{pmatrix}
				0 & \chi^\text{BFISH}_{yyxy} & \chi^\text{BFISH}_{yyxz}\\\chi^\text{BFISH}_{yyyx} & 0 & 0\\\chi^\text{BFISH}_{yzyx} & 0 & 0
			\end{pmatrix}
			\begin{pmatrix}
				0 & \chi^\text{BFISH}_{yzxy} & \chi^\text{BFISH}_{yzxz}\\\chi^\text{BFISH}_{yzyx} & 0 & 0\\\chi^\text{BFISH}_{yzzx} & 0 & 0
			\end{pmatrix}\\
			\begin{pmatrix}
				\chi^\text{BFISH}_{zxxx} & 0 & 0\\0 & \chi^\text{BFISH}_{zyxy} & \chi^\text{BFISH}_{zyxz}\\0 & \chi^\text{BFISH}_{zzxy} & \chi^\text{BFISH}_{zzxz}
			\end{pmatrix}
			\begin{pmatrix}
				0 & \chi^\text{BFISH}_{zyxy} & \chi^\text{BFISH}_{zyxz}\\\chi^\text{BFISH}_{zyyx} & 0 & 0\\\chi^\text{BFISH}_{zzyx} & 0 & 0
			\end{pmatrix}
			\begin{pmatrix}
				0 & \chi^\text{BFISH}_{zzxy} & \chi^\text{BFISH}_{zzxz}\\\chi^\text{BFISH}_{zzyx} & 0 & 0\\\chi^\text{BFISH}_{zzzx} & 0 & 0
			\end{pmatrix}
	\end{pmatrix}}.
	\label{EqBFISHGAFM}
\end{equation}
\\
In our experiment, $\textit{\textbf{B}}=(\cos\beta, \sin\beta, 0)$ is in the $ab$ plane, $\beta$ being the angle between the $\textit{\textbf{B}}$ vector and the crystal $a$ axis. The calculated $P_{\text{spin}}^{2\omega}$ then leads to a prediction of the BFISH pattern $I_\text{spin}^{2\omega}(\varphi)$ (Eq.\,\ref{BFISHIspin}), which is then compared to experimental data. To achieve the best fit, we need to adjust, in principle, the 18 independent tensor elements: $\chi^\text{BFISH}_{xxxy}$, $\chi^\text{BFISH}_{xyxx}$, $\chi^\text{BFISH}_{xzxx}$, $\chi^\text{BFISH}_{xyyy}$, $\chi^\text{BFISH}_{xzyy}$, $\chi^\text{BFISH}_{xzzy}$, $\chi^\text{BFISH}_{yxxx}$, $\chi^\text{BFISH}_{yyxy}$, $\chi^\text{BFISH}_{yzxy}$, $\chi^\text{BFISH}_{yyyx}$, $\chi^\text{BFISH}_{yzyx}$, $\chi^\text{BFISH}_{yzzx}$, 
$\chi^\text{BFISH}_{zxxx}$, $\chi^\text{BFISH}_{zyxy}$, $\chi^\text{BFISH}_{zzxy}$, $\chi^\text{BFISH}_{zyyx}$, $\chi^\text{BFISH}_{zzyx}$, $\chi^\text{BFISH}_{zzzx}$. However, we found that practically, only 5 out of the 18 elements are needed (others can be assumed to be zero); they are $\chi^\text{BFISH}_{xyxx}$, $\chi^\text{BFISH}_{yxxx}$, $\chi^\text{BFISH}_{yyyx}$, $\chi^\text{BFISH}_{xyxx}$, and $\chi^\text{BFISH}_{zzyx}$. The single set of tensor parameters simultaneously fit eight sets of $I_\text{spin}^{2\omega}(\varphi)$ data consisting of four input-output polarization combinations and two $B$ field orientations: $B\parallel a$ ($\beta=0,\pi$) and $B\parallel b$ ($\beta=\pi/2,3\pi/2$). Our best fit is presented in Fig.\,2d(i) of the main text, which is replicated here as Fig.\,\ref{BFISHfitGAFM}. We therefore conclude that the expected surface magnetic point group of the equilibrium intra-layer AFM order can explain the BFISH data.

\subsection{B.~BFISH from metastable magnetic phase}

\begin{figure}[htb]
	\centering
	\includegraphics[width=\linewidth]{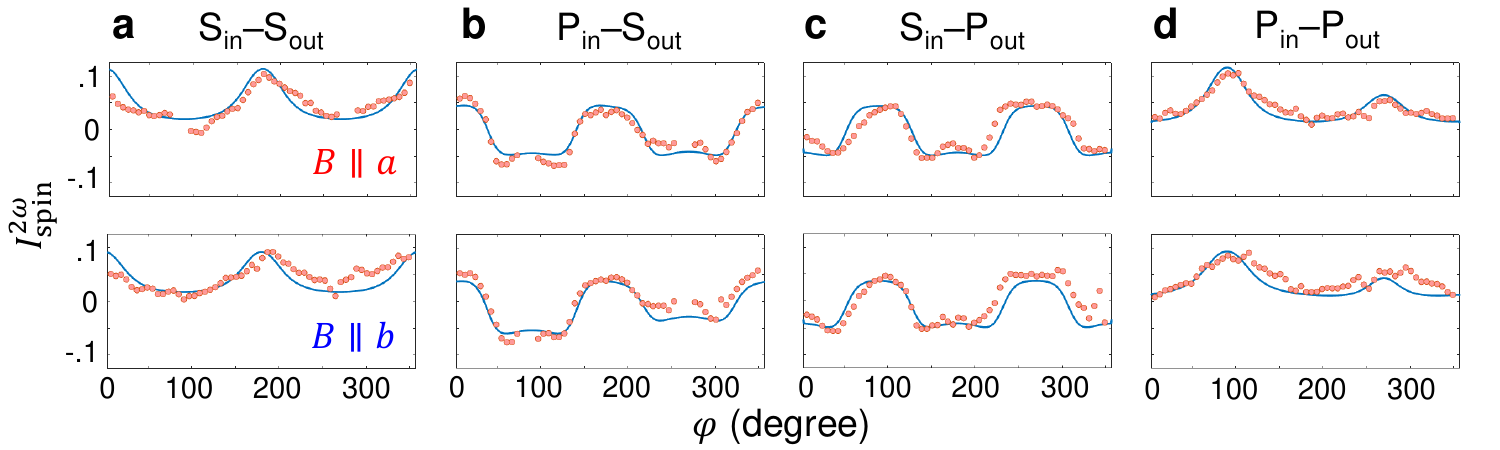}
	\caption{\textbf{Fits to BFISH data in the metastable magnetic phase.} \textbf{a},~S$_\text{in}$--S$_\text{out}$, \textbf{b},~P$_\text{in}$--S$_\text{out}$, \textbf{c},~S$_\text{in}$--P$_\text{out}$, \textbf{d},~P$_\text{in}$--P$_\text{out}$. Top (lower) row panels all correspond to the $B\parallel a$ ($B\parallel b$) configuration. Red circles: experimental data. Blue lines: model calculation.}
	\label{BFISHfitmetastable}
\end{figure}

We carried out the same procedure to fit the BFISH patterns, $I_\text{spin}^{2\omega}(\varphi)$, in the metastable magnetic phase, which is achieved at high pump fluence and at a pump-probe time delay of 10~$\upmu$s. However, we found that if we assume the same surface magnetic group ($m$ with $\sigma_{bc}$) as the intra-layer AFM order, no reasonable fitting can be achieved by tuning the $\chi^\text{BFISH}_{ijkl}$ tensor elements. Key discrepancies occur in the $B\parallel b$ configuration, for which the intra-layer AFM phase shows featureless $I_\text{spin}^{2\omega}(\varphi)$ centered around zero, but the metastable phase shows sizable modulations; see Fig.\,2d(iii) of the main text, which is replicated here as Fig.\,\ref{BFISHfitmetastable}.

To explain the $I_\text{spin}^{2\omega}(\varphi)$ curves in the $B\parallel b$ configuration, we found it necessary to break the $\sigma_{bc}$ glide plane, lowering the surface magnetic group to $1$. This gives the following $\chi_{ijkl}^\text{BFISH}$ tensor
\\
\begin{equation}
	\chi_{ijkl}^\text{BFISH}=
	{\scriptsize \begin{pmatrix}
			\begin{pmatrix}
				\chi^\text{BFISH}_{xxxx} & \chi^\text{BFISH}_{xxxy} & \chi^\text{BFISH}_{xxxz}\\
				\chi^\text{BFISH}_{xyxx} & \chi^\text{BFISH}_{xyxy} & \chi^\text{BFISH}_{xyxz}\\
				\chi^\text{BFISH}_{xzxx} & \chi^\text{BFISH}_{xzxy} & \chi^\text{BFISH}_{xzxz}
			\end{pmatrix}
			\begin{pmatrix}
				\chi^\text{BFISH}_{xyxx} & \chi^\text{BFISH}_{xyxy} & \chi^\text{BFISH}_{xyxz}\\
				\chi^\text{BFISH}_{xyyx} & \chi^\text{BFISH}_{xyyy} & \chi^\text{BFISH}_{xyyz}\\
				\chi^\text{BFISH}_{xzyx} & \chi^\text{BFISH}_{xzyy} & \chi^\text{BFISH}_{xzyz}
			\end{pmatrix}
			\begin{pmatrix}
				\chi^\text{BFISH}_{xzxx} & \chi^\text{BFISH}_{xzxy} & \chi^\text{BFISH}_{xzxz}\\
				\chi^\text{BFISH}_{xzyx} & \chi^\text{BFISH}_{xzyy} & \chi^\text{BFISH}_{xzyz}\\
				\chi^\text{BFISH}_{xzzx} & \chi^\text{BFISH}_{xzzy} & \chi^\text{BFISH}_{xzzz}
			\end{pmatrix}\\
			\begin{pmatrix}
				\chi^\text{BFISH}_{yxxx} & \chi^\text{BFISH}_{yxxy} & \chi^\text{BFISH}_{yxxz}\\
				\chi^\text{BFISH}_{yyxx} & \chi^\text{BFISH}_{yyxy} & \chi^\text{BFISH}_{yyxz}\\
				\chi^\text{BFISH}_{yzxx} & \chi^\text{BFISH}_{yzxy} & \chi^\text{BFISH}_{yzxz}
			\end{pmatrix}
			\begin{pmatrix}
				\chi^\text{BFISH}_{yyxx} & \chi^\text{BFISH}_{yyxy} & \chi^\text{BFISH}_{yyxz}\\
				\chi^\text{BFISH}_{yyyx} & \chi^\text{BFISH}_{yyyy} & \chi^\text{BFISH}_{yyyz}\\
				\chi^\text{BFISH}_{yzyx} & \chi^\text{BFISH}_{yzyy} & \chi^\text{BFISH}_{yzyz}
			\end{pmatrix}
			\begin{pmatrix}
				\chi^\text{BFISH}_{yzxx} & \chi^\text{BFISH}_{yzxy} & \chi^\text{BFISH}_{yzxz}\\
				\chi^\text{BFISH}_{yzyx} & \chi^\text{BFISH}_{yzyy} & \chi^\text{BFISH}_{yzyz}\\
				\chi^\text{BFISH}_{yzzx} & \chi^\text{BFISH}_{yzzy} & \chi^\text{BFISH}_{yzzz}
			\end{pmatrix}\\
			\begin{pmatrix}
				\chi^\text{BFISH}_{zxxx} & \chi^\text{BFISH}_{zxxy} & \chi^\text{BFISH}_{zxxz}\\
				\chi^\text{BFISH}_{zyxx} & \chi^\text{BFISH}_{zyxy} & \chi^\text{BFISH}_{zyxz}\\
				\chi^\text{BFISH}_{zzxx} & \chi^\text{BFISH}_{zzxy} & \chi^\text{BFISH}_{zzxz}
			\end{pmatrix}
			\begin{pmatrix}
				\chi^\text{BFISH}_{zyxx} & \chi^\text{BFISH}_{zyxy} & \chi^\text{BFISH}_{zyxz}\\
				\chi^\text{BFISH}_{zyyx} & \chi^\text{BFISH}_{zyyy} & \chi^\text{BFISH}_{zyyz}\\
				\chi^\text{BFISH}_{zzyx} & \chi^\text{BFISH}_{zzyy} & \chi^\text{BFISH}_{zzyz}
			\end{pmatrix}
			\begin{pmatrix}
				\chi^\text{BFISH}_{zzxx} & \chi^\text{BFISH}_{zzxy} & \chi^\text{BFISH}_{zzxz}\\
				\chi^\text{BFISH}_{zzyx} & \chi^\text{BFISH}_{zzyy} & \chi^\text{BFISH}_{zzyz}\\
				\chi^\text{BFISH}_{zzzx} & \chi^\text{BFISH}_{zzzy} & \chi^\text{BFISH}_{zzzz}
			\end{pmatrix}
	\end{pmatrix}}
\end{equation}
\\
within which, despite having a large number of independent parameters, only 10 of the elements are required to achieve a satisfactory fit (others can be assumed to be zero). These elements are $\chi^\text{BFISH}_{xyxx}$, $\chi^\text{BFISH}_{xyxy}$, $\chi^\text{BFISH}_{yxxx}$, $\chi^\text{BFISH}_{yxxy}$, $\chi^\text{BFISH}_{yyyx}$, $\chi^\text{BFISH}_{yyyy}$, $\chi^\text{BFISH}_{yzyx}$, $\chi^\text{BFISH}_{yzyy}$, $\chi^\text{BFISH}_{zzyx}$, and $\chi^\text{BFISH}_{zzyy}$.
Among these, $\chi^\text{BFISH}_{xyxy}$,  $\chi^\text{BFISH}_{yxxy}$, $\chi^\text{BFISH}_{yyyy}$, $\chi^\text{BFISH}_{yzyy}$, and $\chi^\text{BFISH}_{zzyy}$ has to vanish due to symmetry if the $\sigma_{bc}$ glide plane were to hold; see the tensor structure in Eq.\,\ref{EqBFISHGAFM}. The fact that these elements need to be nonzero to fit the BFISH data in the metastable phase strongly indicates that the $\sigma_{bc}$ glide plane is broken by the metastable magnetic order. The best fits are presented in Fig.\,\ref{BFISHfitmetastable}.

\subsection{C.~Ruling out domain averaging as the origin of the observation}
\begin{figure}[htb]
	\centering
	\includegraphics[width=0.5\linewidth]{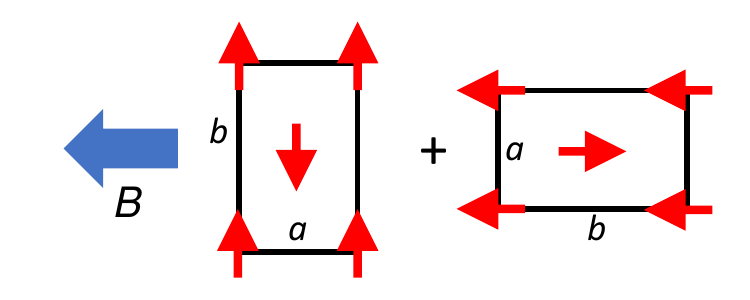}
	\caption{The intra-layer AFM domains under consideration to be averaged out to give the BFISH signal. The rectangles represent orthorhombic structural/magnetic unit cells.}
	\label{Domainavg}
\end{figure}

As shown in Fig.\,\ref{BFISHfitmetastable}, for any particular polarization channel, the $I_\text{spin}^{2\omega}(\varphi)$ lineshapes for $B\parallel b$ and $B\parallel a$ are similar. This is consistent with an intra-layer FM state where moments are aligned perpendicular to both the $a$- and $b$-axes. But it is also important to rule out the possibility that we are simply observing an averaging over pump induced intra-layer FM domains. One scenario, sketched in Fig.\,\ref{Domainavg}, considers a 90$^{\circ}$ domain structure induced by $ab$ twining within the crystal, similar to the one observed in Ca$_3$Ru$_2$O$_7$ \cite{Tang2020}. If the twinning is induced by light, an applied $B$ field will be parallel to $a$ for one domain but parallel to $b$ for the other. This effect, upon averaging over domains within the laser spot, could lead to similar BFISH responses in the $B\parallel b$ and the $B\parallel a$ channels. But this effect can be ruled out for two reasons: (1) The RA-SHG patterns (dominated by the $i$-type contribution) from 90$^{\circ}$ orthorhombic domains are rotated by 90$^{\circ}$ (Fig.\,\ref{Latticepetalfit}a-d) . Therefore a domain averaged RA-SHG pattern would look very different to that from a single domain. Yet we do not observe such a change in our data. (2) As shown in Fig.\,\ref{BFISHfitGAFM}d, the BFISH signals in the P$_\text{in}$--P$_\text{out}$ channel for both $B$ orientations are small in the intra-layer AFM phase. However the curves for the same channel in Fig.\,\ref{BFISHfitmetastable}d show sizable amplitudes of modulations. It is not possible that a weighted average of two small BFISH signals could give a much larger BFISH signal. 

We further considered other domain types such as the 180$^{\circ}$ domains and domains of B-centered intra-layer AFM structure \cite{Braden1998}, but their predictions are at odds with the experimental data as well. For the 180$^{\circ}$ domain, it would preserve the same glide plane that holds in the intra-layer AFM phase, which cannot explain the glide symmetry breaking found in Fig.\,\ref{BFISHfitmetastable}. The B-centered intra-layer AFM domains also cannot explain the light-induced symmetry breaking because they possess the same glide plane as in the A-centered intra-layer AFM phase. Further, one would expect $c$-type Bulk-ED or Bulk-MD to be the dominant radiation channel for BFISH in the B-centered intra-layer AFM phase due to its finite bulk magnetization, but this possibility has been ruled out by the discussions at the beginning of this Section. 

Our data can also rule out other alternative mechanisms that involve magnetic structures sensitive to the orientation of the in-plane magnetic field. One hypothetical scenario corresponds to light-induced weakening of magnetic anisotropy, where in the $F>F_c$ regime the $ab$ plane anisotropy gets weakened to the point where $M$ follows the $B$ direction. Although this picture can explain the appearance of $I_\text{spin}$ signals for both $B\parallel a$ and $B\parallel b$ channels, it is inconsistent with the observed robustness of lattice orthorhombicity against photo-doping, as evidenced by the nearly fluence-independent lattice SHG component at $10\ \mu$s.

\subsection{D.~Summary of magnetic point groups of different phases of \CRO}
\begin{figure}[htb]
	\centering
	\includegraphics[width=0.7\linewidth]{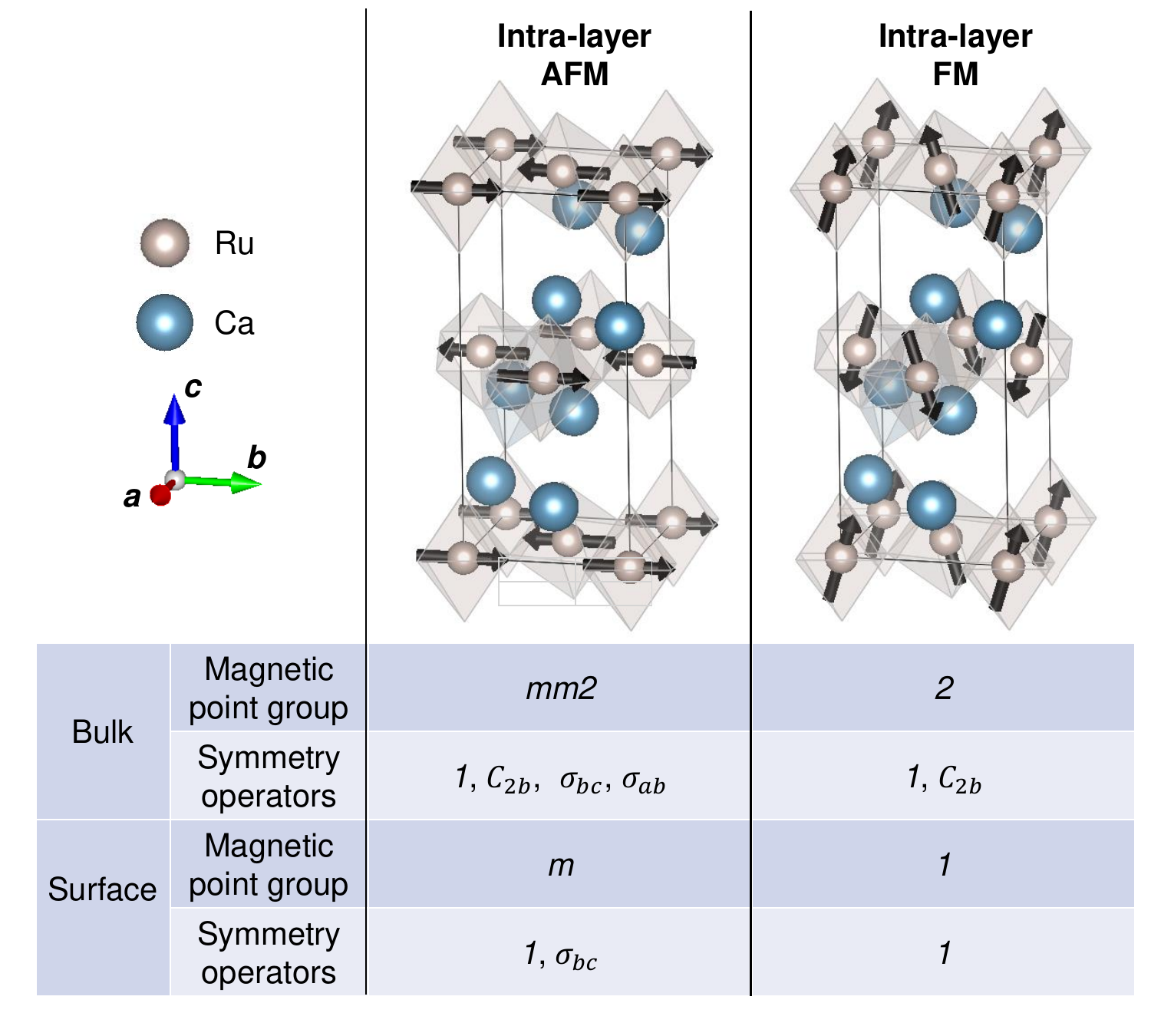}
	\caption{Lattice and spin configurations, and associated magnetic point groups and symmetry operators for \CRO\ in different phases.}
	\label{Scheme_symmetry}
\end{figure}

Figure\,\ref{Scheme_symmetry} presents a summary of magnetic structures and point groups of \CRO\ in different phases. We present this result in terms of point groups because optical measurements in general are insensitive to atomic translation in the space-group operations. Therefore, glide planes would manifest as mirror planes in our measurement.

\newpage

\section{S7.~Optical birefringence as a probe of the lattice symmetry}

\begin{figure}[htb]
	\centering
	\includegraphics[width=0.45\linewidth]{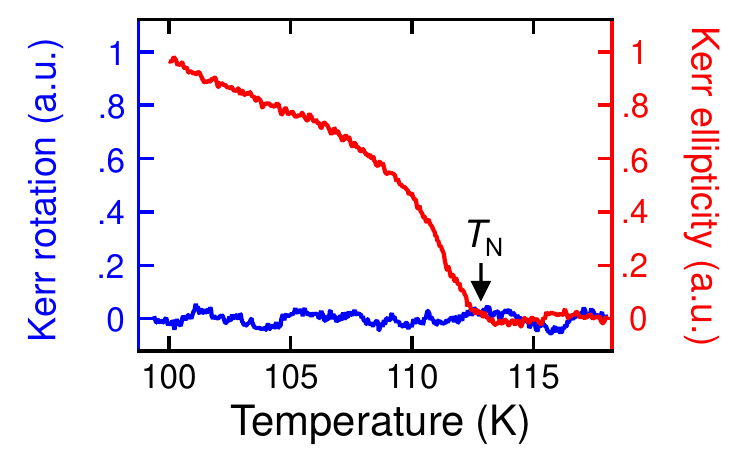}
	\caption{Temperature-dependent Kerr rotation ($F>F_c$ pumping, with a vertical field of 0.15~T) and Kerr ellipticity ($F<F_c$ pumping, no vertical field). Rotation and ellipticity signals are nomalized by the same constant and therefore their magnitudes can be compared.}
	\label{NoKerr}
\end{figure}

Optical birefringence measurements are sensitive to bulk magnetic phase transitions in \CRO. As shown in Fig.\,\ref{NoKerr}, we used a CW laser probe in the reflection geometry to examine polarization changes induced by the crystal. The predominant polarization variation that was observed as a function of temperature was the onset of ellipticity at $T_\text{N}$. No rotation could be observed both in the $F<F_c$ and in the $F<F_c$ regimes. The Kerr ellipticity onset does not depend on if we apply a vertical field ($B=0.15$~T applied along $c$-axis through a permanent ring magnet). The Kerr rotation channel remains silent both with and without the external vertical field. This suggests that optical birefringence is the major contributor of the ellipticity onset, whereas the magneto-optical Kerr effect (in the polar geometry) expected for ferromagnets with bulk magnetization is negligibly small.

\begin{figure}[htb]
	\centering
	\includegraphics[width=0.5\linewidth]{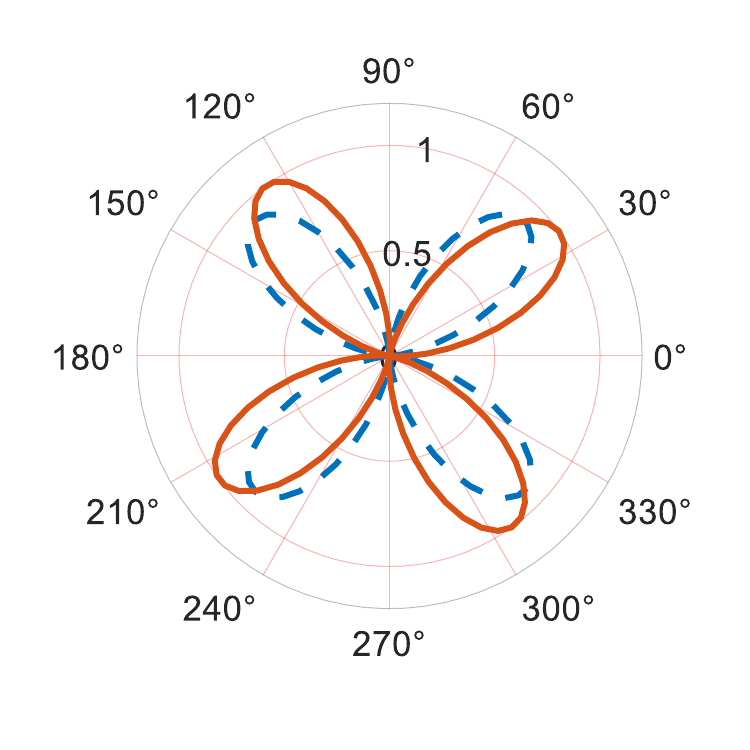}
	\caption{$I^\perp(\varphi)$ signal (normalized) within the cross-polarized S$_\text{in}$--P$_\text{out}$ channel for a crystal with orthorhombic (blue dashed) and monoclinic (red solid) symmetry.}
	\label{Bireforthomono}
\end{figure}

Crystals within the orthorhombic symmetry class exhibit optical birefringence, and in turn, optical birefringence represents a good probe of orthorhombicity. Consider the linear response
\begin{equation}
	P^\omega_i\propto\chi_{ij}E^\omega_j,
\end{equation} 
the polar $i$ tensor $\chi_{ij}$ simplified by generators of the lattice point group $mm2$ would have the structure
\begin{equation}
	\chi_{ij}=
	\begin{pmatrix}
		\chi_{xx} & 0 & 0\\
		0 & \chi_{yy} & 0\\
		0 & 0 & \chi_{zz}
	\end{pmatrix}.
\end{equation}
\\
Therefore, an optical reflectivity measurement in the S$_\text{in}$--P$_\text{out}$ geometry gives
\begin{equation}
	I^\perp(\varphi)\equiv I^\omega_{sp}(\varphi)\propto|P^\omega|^2=|(\chi_{xx}-\chi_{yy})\cos\theta\cos\varphi\sin\varphi|^2
	\label{EqIperp2}
\end{equation}
where $\theta$ and $\varphi$, as defined before, are the angle of incidence and scattering plane angle, respectively. We use $I^\perp$ as the simplified notation of cross-polarized reflection intensity, which is related to the birefringence $\chi_{xx}-\chi_{yy}$. A calculation of $I^\perp(\varphi)$ using Eq.\,\ref{EqIperp2} is shown as the blue dashed line in Fig.\,\ref{Bireforthomono}, which exhibits maxima whenever $\varphi$ bisects the mirror axes ($\varphi=0$ and $\varphi=\pi/2$).

The breaking of the $\sigma_{bc}$ glide plane in the metastable magnetic phase, in addition to exhibiting clear signatures in BFISH described above, can also be captured by $I^\perp(\varphi)$. This is because the glide symmetry breaking lowers the lattice symmetry from orthorhombic to monoclinic. The advantages of measuring $I^\perp(\varphi)$ are its simpler technical implementation and bulk-sensitivity. The disadvantage is that it is less sensitive to symmetry breaking compared to higher-order nonlinear effects like SHG.

When the $\sigma_{bc}$ mirror plane within the $mm2$ ($C_2\parallel b$) point group is broken, the crystal symmetry is lowered to monoclinic point group $2$ ($C_2\parallel b$) or $m$ ($\sigma_{ab}$). Taking $m$ ($\sigma_{ab}$) for example, the $\chi_{ij}$ would be
\begin{equation}
	\chi_{ij}=
	\begin{pmatrix}
		\chi_{xx} & \chi_{xy} & 0\\
		\chi_{xy} & \chi_{yy} & 0\\
		0 & 0 & \chi_{zz}
	\end{pmatrix},
\end{equation}
which gives
\begin{equation}
	I^\perp(\varphi) \propto|P^\omega|^2=|\cos\theta[-\chi_{xy}(\cos\varphi)^2+(\chi_{xx}-\chi_{yy})\cos\varphi\sin\varphi+\chi_{xy}(\sin\varphi)^2]|^2.
	\label{EqIperp}
\end{equation}
An example of the $I^\perp(\varphi)$ lineshape is plotted in Fig.\,\ref{Bireforthomono} as the red curve. Due to the nonzero $\chi_{xy}$ element, the pattern is rotated from the orthorhombic case (blue dashed curve) such that the mirror axes ($\varphi=0$ and $\varphi=\pi/2$) are broken. In addition to pattern rotation, an imbalance of lobe intensities might also occur to break the mirror symmetry, which appears for the point group of $2$ ($C_2\parallel b$). Whichever case appears, the mirror breaking would manifest if one measures $\Delta I^\perp=I^\perp(\varphi_0)-I^\perp(-\varphi_0)$, where $\varphi_0$ is a generic scattering plane angle; this quantity is zero for crystals with orthorhombic symmetry (blue dashed curve in Fig.\,\ref{Bireforthomono}) but nonzero for those with monoclinic symmetry (red solid curve in Fig.\,\ref{Bireforthomono}).

\section{S8.~Modulation-based differential birefringence polarimetry as a probe of glide symmetry breaking} \label{SectrBiref}

Here we provide details of the modulation-based differential birefringence polarimetry technique that provides a measure of the glide-symmetry-breaking magnetic order parameter $\eta_\text{M}$. 

\begin{figure}[htb]
	\centering
	\includegraphics[width=0.5\linewidth]{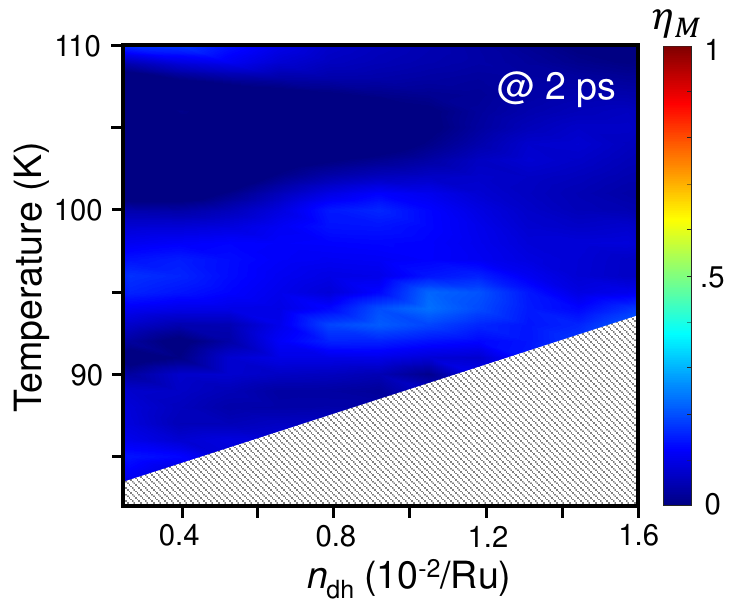}
	\caption{Metastable magnetic order parameter $\eta_\text{M}$ mapped versus temperature and $n_\text{dh}$ at a pump-probe time delay of 2 ps. $\eta_\text{M}=1$ occurs at high fluence and at 10~$\upmu$s (main text Fig.\,2f).}
	\label{NoetaM2ps}
\end{figure}

In a stroboscopic measurement, we have a photodiode reading $I^\perp(\varphi_0)$, and by modulating the pump intensity through an optical chopper, we directly measure the pump-induced change in birefringence $dI^\perp(\varphi_0, t)$ as a function of time delay $t$. We then repeat the measurement at the mirror-symmetric scattering plane angle $-\varphi_0$ and obtain $dI^\perp(-\varphi_0, t)$. Based on the reasoning in the previous section, $\Delta I^\perp=I^\perp(\varphi_0)-I^\perp(-\varphi_0)$ measures the extent of glide symmetry breaking, and therefore, $\Delta I(t) \equiv dI^\perp(\varphi_0,t)-dI^\perp(-\varphi_0,t)$ measures the pump-induced change in the glide-symmetry-breaking order parameter. When the system starts from a glide-symmetry broken state at $t<0$ (equivalent to $t=10~\upmu$s), and the pump pulse restores the glide symmetry, $\Delta I(t>0)$ would be nonzero. On the other hand, if the system is glide symmetric at $t<0$, and the pump-excited state at $t>0$ is still glide symmetric, $\Delta I(t>0)$ would be zero. 

\begin{figure}[htb]
	\centering
	\includegraphics[width=0.9\linewidth]{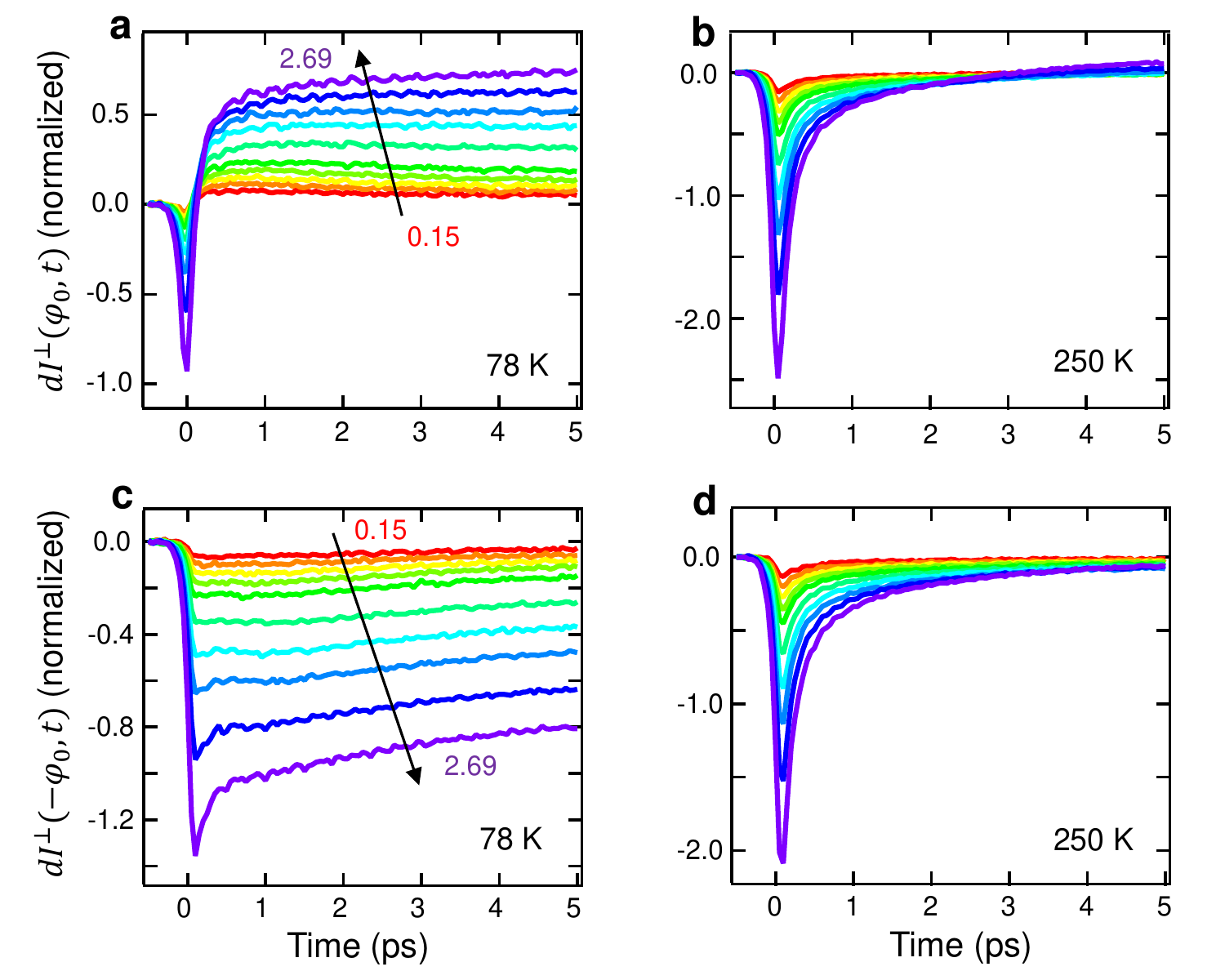}
	\caption{\textbf{a}, $dI^\perp(\varphi_0,t)$ transient (normalized by $\eta_\text{M}$) at 78~K. \textbf{b}, $dI^\perp(\varphi_0,t)$ transient (normalized by $\eta_\text{M}$) at 250~K. \textbf{c}, $dI^\perp(-\varphi_0,t)$ transient (normalized by $\eta_\text{M}$) at 78~K. \textbf{d}, $dI^\perp(-\varphi_0,t)$ transient (normalized by $\eta_\text{M}$) at 250~K. Red to purple: a series of fluences with $n_\text{dh}$=0.15 - 2.69$\times10^{-2}$/Ru.}
	\label{Bireftscans}
\end{figure}

From the time when the pump pulse excites the sample out to $t$ = 100 ps, which is the longest we scanned with our delay stage, the sample is in a glide symmetric state. This is true for any pump fluence and temperature for the following reason. For $F < F_c$, the system remains in the glide-symmetricintra-layer AFM state and therefore the metastable state is not accessed. For $F > F_c$, even though the metastable state appears at late times, the system is in a glide-symmetric paramagnetic state at these early times. This is corroborated by BFISH measurements of $\eta_\text{M}$ at 2~ps (Fig.\,\ref{NoetaM2ps}), which shows an absence of $\eta_\text{M}$ across all measured fluences and temperatures. Taking this argument into account, it is clear that $\Delta I(t)$ provides a measure of the glide-symmetry-breaking order parameter $\eta_\text{M}$ at $t < 0$ (i.e. 10~$\upmu$s). This is the same quantity that could have been measured by $\Delta I^\perp=I^\perp(\varphi_0)-I^\perp(-\varphi_0)$, but $\Delta I(t)$ obtained through the modulation spectroscopy technique has much higher sensitivity that facilitates detection of small signals.

\begin{figure}[htb]
	\centering
	\includegraphics[width=0.9\linewidth]{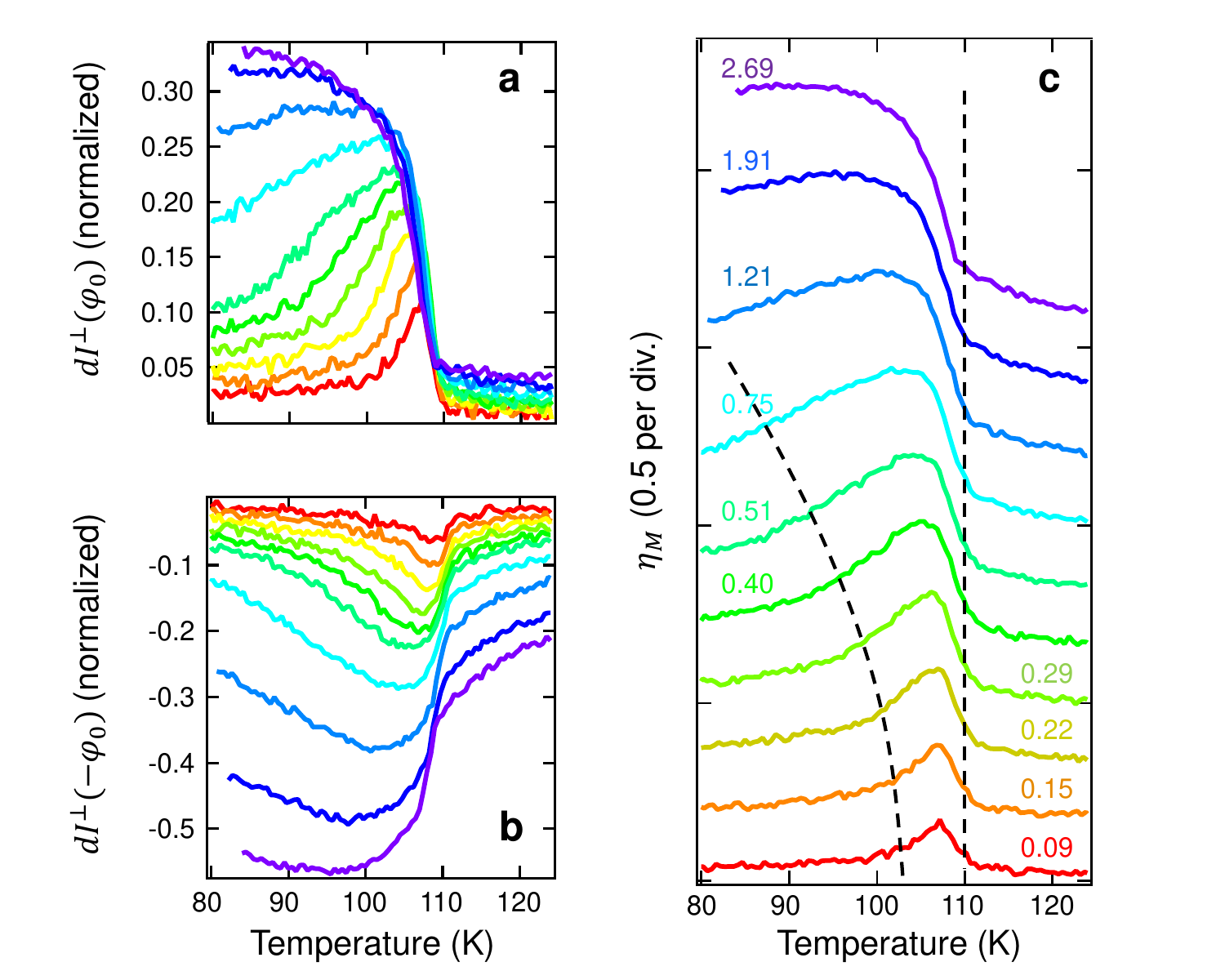}
	\caption{\textbf{a}, $dI^\perp(\varphi_0)$ (normalized by $\eta_\text{M}$) versus temperature at 50~ps. \textbf{b}, Same as \textbf{a} but plotting $dI^\perp(-\varphi_0)$. \textbf{c}, $\eta_\text{M}$ (calculated from $dI^\perp(\varphi_0)-dI^\perp(-\varphi_0)$) versus temperature. Red to purple: a series of fluences with $n_\text{dh}$=0.09 - 2.69$\times10^{-2}$/Ru. Curves are offset. Dashed black lines are phase boundaries, where the region between them is the metastable magnetic phase with large $\eta_\text{M}$. This phase diagram resembles a 90$^\circ$ rotated version of Fig.\,3c(i) of the main text. Color legend of fluence for \textbf{a} and \textbf{b} is the same as that for \textbf{c}.}
	\label{BirefTDep}
\end{figure}

To further prove this idea, we plot $dI^\perp(\varphi_0,t)$ and $dI^\perp(-\varphi_0,t)$ both below and well above $T_\text{N}$ (Fig.\,\ref{Bireftscans}). There is a clear difference between $dI^\perp(\varphi_0,t)$ and $dI^\perp(-\varphi_0,t)$ at 78~K (thereby implying a large $\Delta I(t)$ and $\eta_\text{M}$ for strong pump fluences), but no difference exists at 250~K where metastable order is absent.

Figure~3b and 3c of the main text are obtained by plotting $\Delta I(t)$ as a function of fluence at different temperatures. Here we supplement this data with temperature dependent scans of $\Delta I(t)$ at various pump fluences, which too can reconstruct the $\eta_\text{M}$ phase diagram. This data set is shown in Fig.\,\ref{BirefTDep}, where again from panels a and b we see a large discrepancy between $dI^\perp(\varphi_0,t)$ and $dI^\perp(-\varphi_0,t)$. The temperature dependent $\eta_\text{M}$ [measured by taking $\Delta I(t)$] at various fluences clearly show a region (that bounded between the two black dashed lines in panel c) that can be assigned to the metastable order. As expected, these phase boundaries closely match those shown in Figure 3(c)iii of the main text. 

\section{S9.~Two-site exact diagonalization model} \label{Sectwosite}

We constructed a two-site exact diagonalization model to capture the potential mechanism for the magnetic phase transition. The key observation from this theory is a intra-layer AFM to intra-layer FM transition as the structural parameter $\Delta$ is reduced (Fig.\,4 of main text). We first provide some intuition for this prediction and then illustrate the details of our model. 

\begin{figure}[htb]
	\centering
	\includegraphics[width=0.9\linewidth]{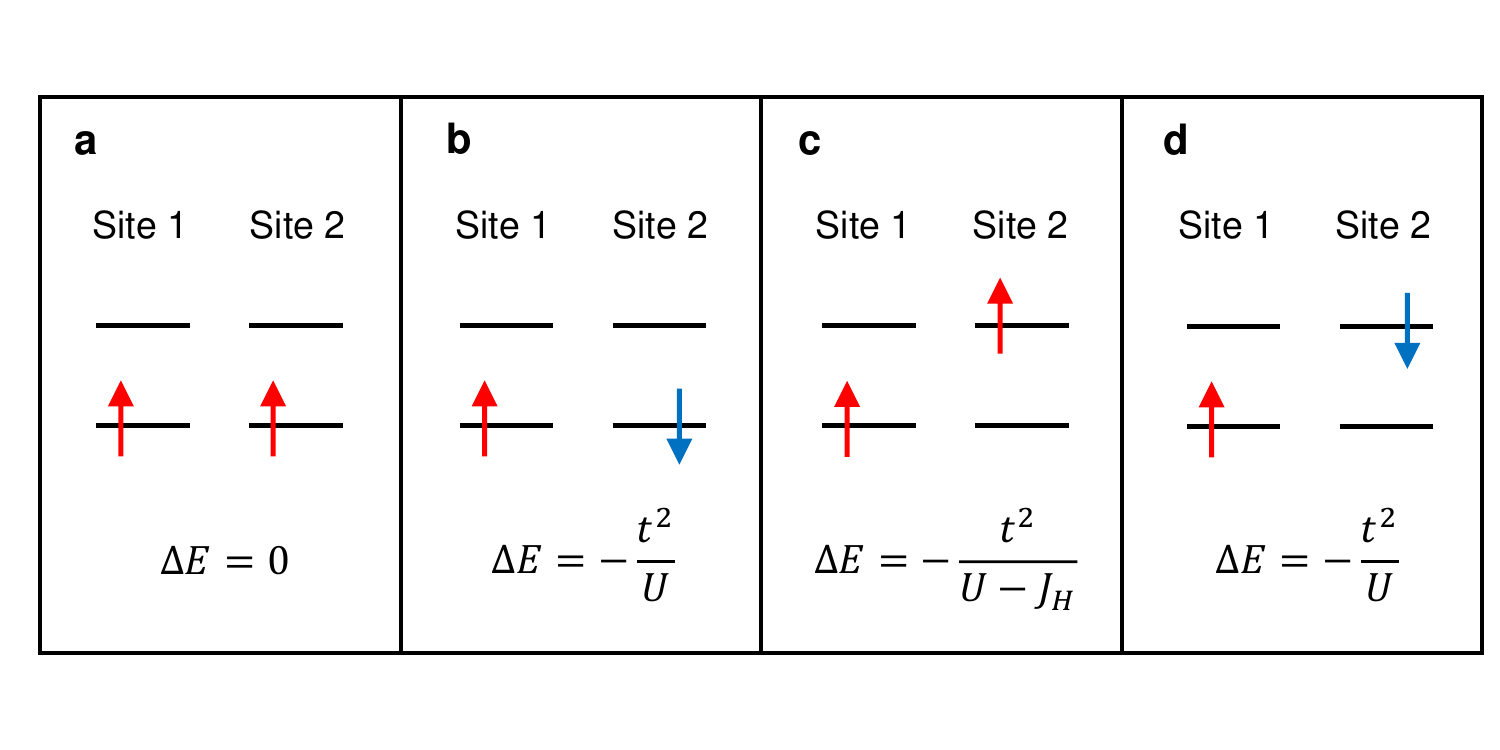}
	\caption{\textbf{Two-site degenerate two-orbital, quarter-filled Hubbard model.} \textbf{a}, the ferro-orbital ferromagnetic configuration. \textbf{b}, the ferro-orbital antiferromagnetic configuration. \textbf{c}, the antiferro-orbital ferromagnetic configuration. \textbf{d}, the antiferro-orbital antiferromagnetic configuration. Consider the two orbitals being energetically degenerate, the atomic energies of these states are labeled beneath the graphs.}
	\label{Quarterfill}
\end{figure}

The most direct intuition comes from the multi-orbital Mott physics \cite{Khomskii2022}. Consider a two-site, two-orbital (energetically degenerate), quarter-filled system as shown in Fig.\,\ref{Quarterfill}. Strong Coulomb repulsion forces the two electrons to occupy different sites, and four major types of low-energy orbital and spin arrangements can be identified. They are (1) the ferro-orbital ferromagnetic configuration (Fig.\,\ref{Quarterfill}a), (2) the ferro-orbital antiferromagnetic configuration (Fig.\,\ref{Quarterfill}b), (3) the antiferro-orbital ferromagnetic configuration (Fig.\,\ref{Quarterfill}c), and (4) the antiferro-orbital antiferromagnetic configuration (Fig.\,\ref{Quarterfill}d). Among these, due to the virtual inter-site charge hopping, the ferro-orbital antiferromagnetic configuration and the antiferro-orbital antiferromagnetic configuration have an additional energy saving of $\Delta E=-t^2/U$, where $t$ denotes the spin- and orbital-conserved inter-site hopping amplitude. Notably, the antiferro-orbital ferromagnetic configuration has the lowest energy, with $\Delta E=-t^2/(U-J_\text{H})$, because not only is the virtual charge hopping is active, but the spin alignment after hopping is energetically favorable by Hund's coupling ($J_\text{H}$) as well. This conclusion provides the general intuition that for degenerate multi-orbital models away from half filling, Hund's coupling favors ferromagnetic spin-spin correlations.

Now consider breaking orbital degeneracy in the model above. When the orbital energy gap $\Delta$ is increased, the system favors occupying one orbital over the other, which tends to have ferro-orbital configuration. This turns off the hopping pathway from filled to empty orbitals. The ground-state solution should instead be the ferro-orbital antiferromagnetic configuration (Fig.\,\ref{Quarterfill}b). A magnetic phase transition induced by tuning $\Delta$ is therefore expected.

Although \CRO\ has a different orbital and filling configuration as the toy model above, the essential physics of the $\Delta$-tuned magnetic phase transition is similar. When $\Delta=0$, the system features a multi-orbital configuration away from half filling, and when $\Delta$ is large, the system features ferro-orbital ordering with half-filled $d_{yz/xz}$ orbitals. Following the treatment in Ref.\,\cite{Meetei2015}, a two-site model for \CRO\ was established to study the impact of $\Delta$ on the magnetic order. The Hamiltonian is
\\
\begin{equation}
	H = H_\text{hop} + H_\text{e-latt} + H_\text{int} + H_\text{SOC},
	\label{Htot}
\end{equation}
where
\begin{align}
	H_\text{hop} & = \sum_{ij,\alpha,\sigma}(t^\alpha_{ij}c^\dagger_{i\alpha\sigma}c_{j\alpha\sigma}+\text{H.c.}), \\
	H_\text{e-latt} & = -\Delta\sum_{i,\sigma}[n_{ixy\sigma}-\frac{1}{2}(n_{iyz\sigma}+n_{ixz\sigma})],\\
	H_\text{int} & = U\sum_{i,\alpha,\sigma_1\neq\sigma_2}n_{i\alpha\sigma_1}n_{i\alpha\sigma_2}+(U-2J_\text{H})\sum_{i,\alpha\neq\beta,\sigma_1\sigma_2}n_{i\alpha\sigma_1}n_{i\beta\sigma_2}\nonumber\\& -J_\text{H}\sum_{i,\alpha\neq\beta,\sigma_1\sigma_2}c^\dagger_{i\alpha\sigma_1}c_{i\alpha\sigma_2}c^\dagger_{i\beta\sigma_2}c_{i\beta\sigma_1},\\
	H_\text{SOC} & = \lambda\sum_{ij,\alpha\beta,\sigma_1\sigma_2}\braket{\vb{s}\cdot\vb{l}}_{\alpha\sigma_1,\beta\sigma_2}c^\dagger_{i\alpha\sigma_1}c_{i\beta\sigma_2},
	\label{twositeH}
\end{align}
\\
are, respectively, the charge hopping, electron-lattice coupling, Kanamori interaction, and spin-orbit coupling terms. $i$, $j$ are site indices, $\alpha,\beta\in\{xy,yz,xz\}$ are orbital indices, and $\sigma_1,\sigma_2\in\{\uparrow,\downarrow\}$ are spin indices. $c^\dagger$ and $n=c^\dagger c$ are the electron creation operator and number operator, respectively. $t^\alpha_{ij}$, $U$, $J_\text{H}$, and $\lambda$ are the orbital-dependent hopping amplitude, on-site Coulomb interaction, Hund's coupling, and spin-orbit coupling parameters, respectively. Compared to the mean-field model detailed in Section~S2, this model is solved by exact diagonalization, which is able to include the contribution of orbital moments, spin-orbit coupling, and off-diagonal charge condensates (off-diagonal terms of $\langle c^\dagger_{i\,\alpha\,\sigma_1}c_{j\,\beta\,\sigma_2}\rangle$) \cite{Mohapatra2020}; these terms were dropped by the Hartree-Fock approximation in the mean-field model.

To calculate the magnetic ground state as in Fig.\,4c of the main text, we set the hopping amplitudes to realistic anisotropic values: $t_{xy}=1.5$, $t_{yz}=1$, $t_{xz}=0$, and $U=10$, $J_\text{H}=2.5$, $\lambda=0.4$, and varied $\Delta$. Here, we did not adopt the assumption of isotropic diagonal hopping, because in the realistic case, due to the one-dimensional nature of $d_{yz}$ and $d_{xz}$ orbitals, $t_{yz}$ and $t_{xz}$ should be different for any nearest-neighbor Ru ion pair. However, just as what has been found in Ref.\,\cite{Meetei2015}, the essential predictions originating from the model are qualitatively similar to the isotropic case, which we have confirmed by our simulations.

To calculate the magnetic anisotropy in the metastable FM phase (Fig.\,4d of the main text), we fixed $\Delta=0.3$, a value for which the ground-state solution has FM correlations (Fig.\,4c of the main text). Then, using the expression for the second-order anisotropy energy $E_A=K_{ac}[\sin(\theta)]^2$, the anisotropy constant $K_{ac}$ was calculated using the idea of torque magnetometry. We set up a weak magnetic field $\textit{\textbf{B}}$ in our calculation, and scanned the angle of $\textit{\textbf{B}}$ with respect to the $a$ axis within the $ac$ plane. This will cause the net magnetization vector $\textit{\textbf{M}}$ to rotate within the plane. Due to the anisotropy energy $E_A$, $\textit{\textbf{M}}$ will not be exactly parallel to $\textit{\textbf{B}}$, and one can apply the expression from torque magnetometry
\begin{equation}
	K_{ac}\sin(2\theta)= |\textit{\textbf{M}}\times\textit{\textbf{B}}|,
\end{equation}
where $\theta$ denotes the angle between $\textit{\textbf{M}}$ and the $a$ axis. Finding the maximum value of $|\textit{\textbf{M}}\times\textit{\textbf{B}}|$ by scanning $\theta$ therefore allows us to determine $K_{ac}$. The calculated result is plotted in Fig.\,4d of the main text.

\begin{figure}[htb]
	\centering
	\includegraphics[width=0.8\linewidth]{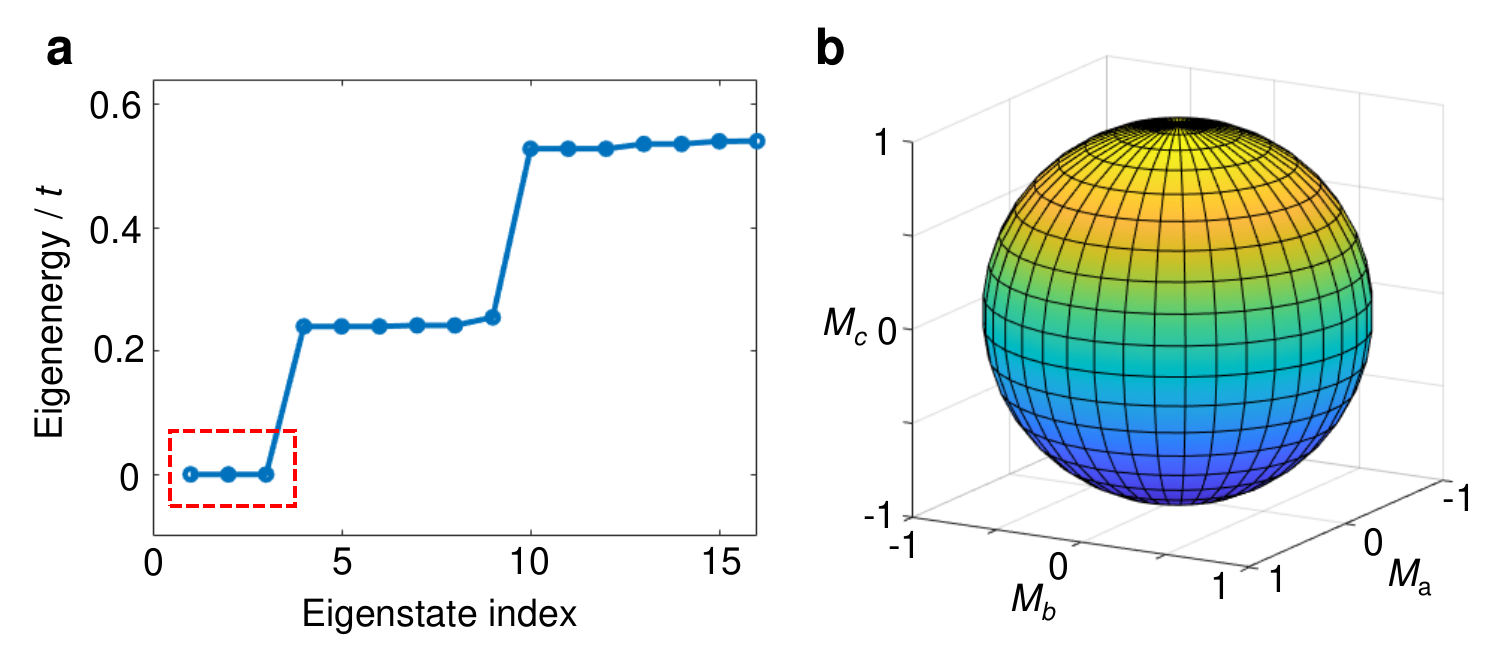}
	\caption{\textbf{Isotropic limit of the intra-layer FM state revealed by the two-site model.} \textbf{a}, Lowest eigenstates and corresponding eigen-energies. \textbf{b}, $\bm{M}\cdot \bm{B}$ mapped with a weak $\bm{B}$ field scanned throughout the entire solid angle. Induced $\bm{M}$ is isotropic and always directed along $\bm{B}$.}
	\label{IsotropicFM}	
\end{figure}

Finally, we provide some intuition to the weak magnetic anistropy in the $ac$ plane. For this purpose, we identified a benchmarking limit that is known to give isotropic intra-layer FM order. With other parameters taking realistic values, we found this limit occurs when (1) orbital order vanishes ($\Delta=0$) and (2) hopping amplitudes are isotropic $t_{xy}=t_{yz}=t_{xz}=t$. Our calculation clearly shows three energetically degenerate ground states (Fig.\,\ref{IsotropicFM}), reminiscent of the spin-triplet ground state representing isotropic FM order in a half-filled two-site Mott-Hubbard model \cite{Mentink2015}. Further, with a small polarizing $\bm{B}$ field in the model, we found $\bm{M}$ in the ground state maintains constant magnitude and always points in the same direction as $\bm{B}$ (Fig.\,\ref{IsotropicFM}). Such an isotropic model of \CRO\ is also discussed in Ref.~\cite{Meetei2015} which, upon mapping to an effective spin Hamiltonian (Eq.~6 in Ref. \cite{Meetei2015}), respects full rotational symmetry.

Now we discuss the two discrepancies between the parameter set used in our realistic model and the isotropic benchmark, and analyze their impacts to the magnetic anisotropy.

(1) $\Delta\neq0$: Finite $\Delta$ lifts the $t_{2g}$ orbital degeneracy and further lowers the symmetry of the crystal-field environment, leading to anisotropy. However, we believe its impact is small. This is particularly true for the intra-layer FM state because, due to the cooperative spin and orbital response across the transition, the $\Delta$ tied to the intra-layer FM state is smaller than that of the intra-layer AFM ground state (See schematic in Fig.\,1a and the $\Delta$-dependent energy landscape in Fig.\,4c). In the realistic case, the Hund’s coupling energy is a few times larger than $\Delta$ for the intra-FM phase, which equalizes the electron population among the $t_{2g}$ orbitals, thereby minimizing the effect of $\Delta$.

(2) Anisotropic hopping amplitudes: Within a corner-shared octahedral network, the orbital-dependent hopping amplitudes ought to be anisotropic. This effect, however, has been fully addressed by prior theoretical studies \cite{Meetei2015}. Appendix D of Ref.~\cite{Meetei2015} compares the low-energy solutions of the two-site model of \CRO\ adopting isotropic and anisotropic hopping amplitudes. Even assuming highly anisotropic hopping, the calculation shows that the triplet state marking the $J = 1$ FM state only becomes slightly nondegenerate, which is consistent with weak magnetic anisotropy.

Taken together, we conclude that the deviation of our model parameters in the intra-layer FM phase from the isotropic benchmark is not significant enough to give strong magnetic anisotropy.

\section{S10.~Absence of light-induced insulator-to-metal transition}

\begin{figure}[htb]
	\centering
	\includegraphics[width=0.95\linewidth]{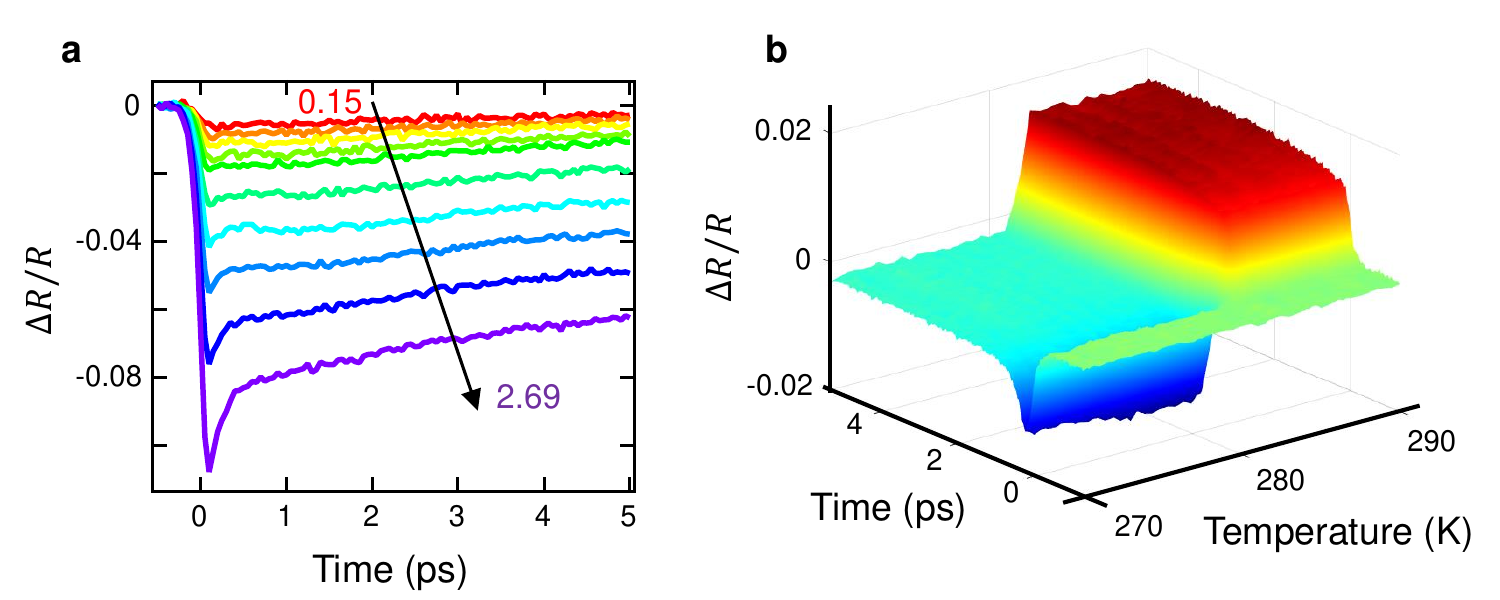}
	\caption{\textbf{a}, Fluence dependent differential reflectivity transients at 78~K. Red to purple: a series of fluences with $n_\text{dh}$=0.15 - 2.69$\times10^{-2}$/Ru. The highest fluence is much larger than the critical fluence of the metastable phase transition. \textbf{b}, Differential reflectivity transients across a temperature-induced insulator-to-metal transition in \CRO. The crystal was slightly doped by Cr to lower the transition temperature to around room temperature.}
	\label{dRRIMT}
\end{figure}

Although the light-induced metastable phase has a cooperative spin-orbital character, this phase remains insulting up to the highest pump fluence measured. We reached this conclusion by comparing the fluence-dependent differential reflectivity ($\Delta R/R$) traces at base temperature (Fig.\,\ref{dRRIMT}a) with the traces across a temperature-driven insulator-to-metal (IMT) phase transition.

\CRO\ shows a pronounced Mott IMT at a temperature higher than the orbital ordering temperature. By measuring the temperature-dependent $\Delta R/R$ at a fixed fluence, we observed a pronounced sign change of the signal across IMT (Fig.\,\ref{dRRIMT}b). The sample we used for this test was a Cr-doped \CRO, whose IMT temperature is lowered to near room temperature, which eases experimental implementation \cite{Qi2010}. Up to the largest pump fluence $n_\text{dh}$=2.69$\times10^{-2}$/Ru (the critical fluence for inducing the metastable order is $n_\text{dh}$=0.8$\times10^{-2}$/Ru), no such sign change in $\Delta R/R$ was observed in the pristine \CRO\ crystal. 

\begin{figure}[htb]
	\centering
	\includegraphics[width=0.45\linewidth]{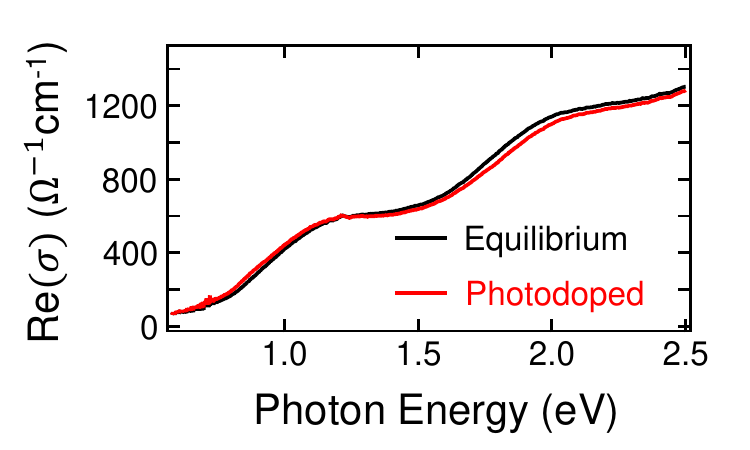}
	\caption{Optical conductivity spectra of \CRO\ in the un-pumped equilibrium state at 80 K and the 1 eV pumped non-equilibrium state at a time delay of 0.5 ps ($F=4$ mJ/cm$^2$, $n_\text{dh}=3.8\times10^{-2}$/Ru).}
	\label{dsigma}
\end{figure}

Absence of a light-induced IMT is also supported by our prior pump-probe optical conductivity measurements \cite{Li2022}. Figure~\ref{dsigma} shows the broadband optical conductivity spectrum of \CRO\ at 80~K. In equilibrium, the spectrum features an insulating gap of around 0.6~eV, and two peak structures at 1.2~eV and 2~eV, which are attributed to the  $d_{xy}\rightarrow d_{yz/xz}$ and $ d_{yz/xz}\rightarrow d_{yz/xz}$ transitions, respectively. Maintaining the same conditions, we photodoped the crystal with 1~eV pump pulses at a fluence of $F=4$ mJ/cm$^2$ ($n_\text{dh}=3.8\times10^{-2}$/Ru), far higher than $F_c$ for the metastable transition. The nonequilibrium conductivity spectrum at a time delay of 0.5~ps is overlaid in Fig.\,\ref{dsigma} as the red curve. There is minimal spectral weight transfer and the gap clearly remains intact. Moreover, recent work on epitaxially grown \CRO\ films \cite{Verma2024} showed a photo-induced IMT occurring above a fluence threshold larger by at least an order of magnitude compared to $F_c$ of our study.

\section{S11.~Presence of metastable magnetic domains}

\begin{figure}[htb]
	\centering
	\includegraphics[width=0.55\linewidth]{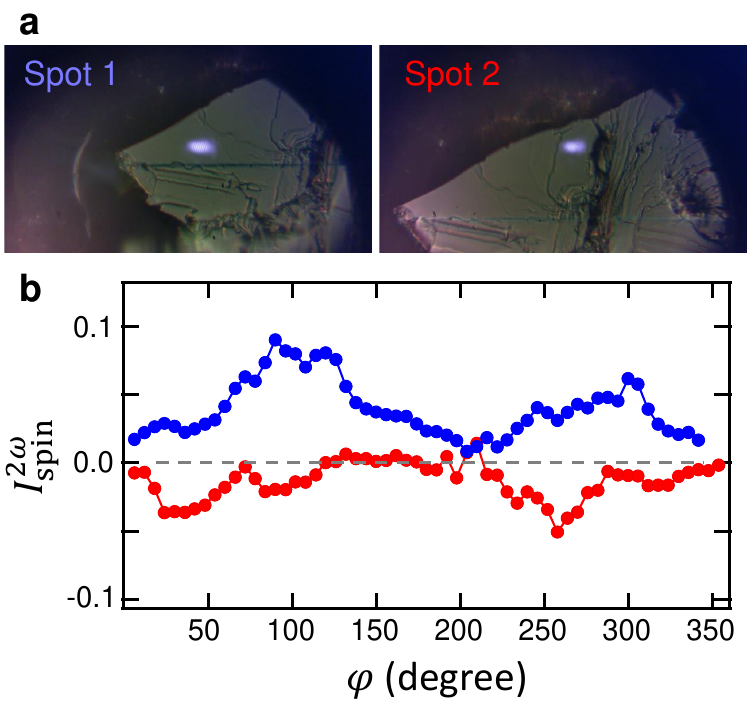}
	\caption{\textbf{a}, Laser spot sampling different locations of the crystal. \textbf{b}, P$_\text{in}$--P$_\text{out}$ $I_\text{spin}^{2\omega}$ patterns  collected at spot 1 and 2 (in \textbf{a}) in the light-induced metastable state (at high fluence and low temperature); data from spot 1 is reported in the main text. The opposite phase of the intensity modulations suggest opposite metastable magnetic order parameters.}
	\label{FMdomain}
\end{figure}

Since our symmetry analysis and theoretical model suggest that the metastable order has a FM character, one natural question that arises is whether domains could exist across different spatial locations of the crystal. We tested this by taking BFISH measurements at a few locations on the crystal surface. Two representative spots we sampled are shown in Fig.\,\ref{FMdomain}a, where the bright purple dot on the whitelight image is the focused laser spot.

Figure~\ref{FMdomain}b shows the $I_\text{spin}^{2\omega}$ signals we collected through BFISH measurements in the P$_\text{in}$--P$_\text{out}$ channel on the two spots at 78~K and under strong pumping. The fact that both traces show intensity modulations that deviate from zero suggests that strong fluence can induce metastable order at both spots. Notably, the intensity modulations carry opposite signs, suggesting that the metastable magnetic order parameters are also opposite in sign between the two locations. The two spots therefore belong to different domain states of the light-induced order.

We note that since the magnetic order parameter is captured by stroboscopic measurements that average over many laser pulse cycles, the domain state of the metastable order appears to be pinned from pulse to pulse at a particular spatial location. Although the mechanism is not entirely clear, we speculate that local structural distortions may be at play, leading to a pinning of magnetic domain orientations.

\section{S12.~Persistent effects in stroboscopic experiments}

For pumping conditions strong enough to induce the intra-layer FM state, the system’s relaxation dynamics takes longer than the laser pulse period. This means that, in our stroboscopic measurements, we are probing the photo-excited sample whose properties do not fully recover to equilibrium between the repetitive pump pulses. Below we address the important question on model compatibility related to this experimental scenario.

Our model in Fig.\,4e most ideally applies to the dynamics triggered by the first pump pulse, which initiates from the intra-layer AFM and ends with the intra-layer FM state. For subsequent pulses, the dynamics would start from the intra-layer FM state. However, we note that our model is still highly relevant, especially within the 1 ps - 10 µs window for every pulse cycle, during which the metastable intra-layer FM order emerges from a hot transient nonmagnetic state. Regardless of the initial state, the system’s status at step ii (around 1 ps after a strong pump) is always a hot transient nonmagnetic state. The ensuing evolution of the energy landscape and population dynamics after this time snapshot, which represents the most crucial trapping mechanism of the time-hidden metastable state reported by this work, is consistent throughout all pulses and well captured by our model.

The data shown in Fig. 1e of the main text were acquired in the standard stroboscopic pump-probe geometry where RA SHG signals at a given time delay are averaged over many pump-probe cycles. Since the pump fluence used in Fig.\,1e ($n_\text{dh}=1.9\times10^{-2}$/Ru) was above the threshold for inducing the metastable state, the pre-time-zero snapshot is in the metastable intra-layer FM state. This is one scenario where the persistent effect can impact the interpretation of our experiments. However, we have repeated the same measurement with a below-threshold pump fluence ($n_\text{dh}=0.6\times10^{-2}$/Ru) where the dynamics start from the equilibrium intra-layer AFM state for every pulse cycle. We observed the same photo-doping induced reduction of $\Delta$ (shown below in Fig.\,\ref{BeforeAfterIlatt}).

\begin{figure}[htb]
	\centering
	\includegraphics[width=0.45\linewidth]{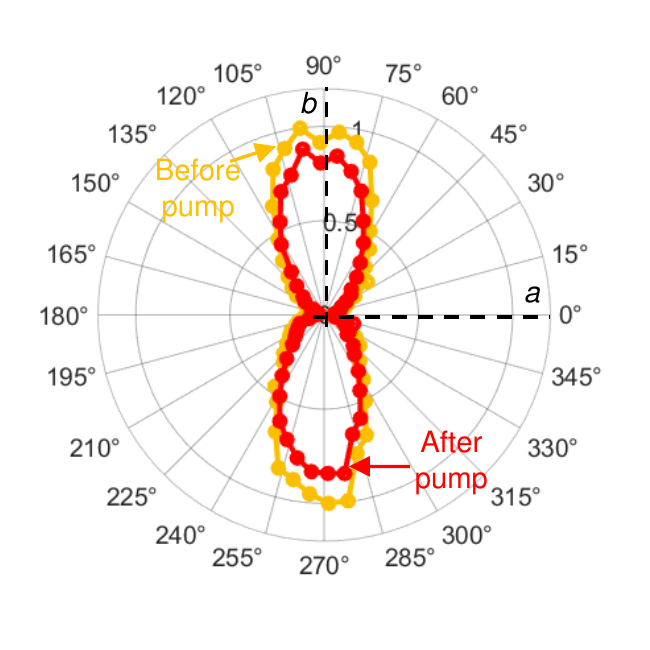}
	\caption{$I^{2\omega}(\varphi)$ plots for S$_\text{in}$--P$_\text{out}$ geometry before (orange) and 0.1~ps after (red) impulsive photo-doping at 80 K. Fluence: $n_\text{dh}=0.6\times10^{-2}$/Ru.}
	\label{BeforeAfterIlatt}
\end{figure}

\newpage


\end{document}